\pdfoutput=1
\documentclass[a4paper,12pt]{article}
\usepackage[hmargin=2cm,vmargin=2cm]{geometry}
\usepackage{amsthm}
\usepackage{amsmath}
\usepackage{amssymb}
\usepackage{enumitem}
\usepackage{array}
\usepackage{authblk}
\usepackage{algorithm}
\usepackage{rotating}
\usepackage{algpseudocode}
\usepackage[compact]{titlesec}
\usepackage{graphicx}
\usepackage{caption}
\usepackage[mathscr]{eucal}
\usepackage[pdftex]{color}
\usepackage[text]{datetime}
\definecolor{darkblue}{rgb}{0.7,0.2,0.2}
\definecolor{lightred}{rgb}{1,0.2,0.6}
\usepackage[orig, british, cleanlook]{isodate}
\definecolor{linkcol}{rgb}{0.9,0.2,0.1}
\usepackage[pdftitle={},pdfauthor={},colorlinks=TRUE,allcolors=linkcol]{hyperref}
\usepackage[backend=bibtex, natbib=true, bibstyle=numeric, isbn=false, citestyle=numeric, doi=true, date=year, sorting=nyt, giveninits, maxbibnames=50]{biblatex}
\renewbibmacro{in:}{}
\AtEveryBibitem{%
    \clearlist{language}%
}
\newcommand{\ud}{\mathrm{d}}

\newcommand{\eqdefr}{=\mathrel{\mathop:}}
\newcommand{\eqdefl}{\mathrel{\mathop:}=}
\newcommand{\N}{\mathbb{N}}

\newcommand{\R}{\mathbb{R}}

\newcommand{\E}{\mathbb{E}}

\newcommand{\prob}{\mathbb{P}}

\theoremstyle{plain}

\theoremstyle{definition}

\theoremstyle{remark}
\newtheorem{rem}[equation]{Remark}
\newtheorem{exm}[equation]{Example}

\begin{document}

\title{Unifying incidence and prevalence under a time-varying general branching process}
\date{\today}

\author[1,2,$\P$,*]{Mikko S. Pakkanen}
\author[2,*]{Xenia Miscouridou}
\author[3,*]{Matthew J. Penn}
\author[4]{\authorcr Charles Whittaker}
\author[2]{Tresnia Berah}
\author[4,5]{ Swapnil Mishra}
\author[4,*]{\authorcr Thomas A. Mellan} 
\author[4,5,$\P$,*]{Samir Bhatt}

\affil[$1$]{Department of Statistics and Actuarial Science, University of Waterloo}
\affil[$2$]{Department of Mathematics, Imperial College London}
\affil[$3$]{Department of Statistics, University of Oxford}
\affil[$4$]{Department of Infectious Disease Epidemiology, Imperial College London}
\affil[$5$]{Section of Epidemiology, Department of Public Health, University of Copenhagen}

\affil[$*$]{Joint authorship}
\affil[$\P$]{Corresponding authors: \href{mailto:mikko.pakkanen@uwaterloo.ca}{\texttt{mikko.pakkanen@uwaterloo.ca}}, \href{mailto:s.bhatt@imperial.ac.uk}{\texttt{s.bhatt@imperial.ac.uk}}}

\renewcommand\Authsep{, }
\renewcommand\Authands{ and }
\renewcommand\Affilfont{\small}

\maketitle
\begin{abstract}
Renewal equations are a popular approach used in modelling the number of new infections, i.e., incidence, in an outbreak. We develop a stochastic model of an outbreak based on a time-varying variant of the Crump--Mode--Jagers branching process. This model accommodates a time-varying reproduction number and a time-varying distribution for the generation interval. We then derive renewal-like integral equations for incidence, cumulative incidence and prevalence under this model. We show that the equations for incidence and prevalence are consistent with the so-called back-calculation relationship. We analyse two particular cases of these integral equations, one that arises from a Bellman--Harris process and one that arises from an inhomogeneous Poisson process model of transmission. We also  show that the incidence integral equations that arise from both of these specific models agree with the renewal equation used ubiquitously in infectious disease modelling. We present a numerical discretisation scheme to solve these equations, and use this scheme to estimate rates of transmission from serological prevalence of SARS-CoV-2 in the UK and historical incidence data on Influenza, Measles, SARS and Smallpox.

\vspace*{1em}

\end{abstract}

\section{Introduction}

Mathematical descriptions of infectious disease outbreaks are fundamental to forecasting and simulating the dynamics of epidemics, as well as to understanding the mechanics of how transmission occurs. Epidemiological quantities of interest include \emph{incidence} (the number of new infections at a given time point), \emph{cumulative incidence} (the total number of infections up to a given time point) and \emph{prevalence} (the number of infected individuals at a given time point). Taking a somewhat reductive perspective, it can be said that two main popular frameworks co-exist when modelling an infectious disease outbreak, namely, \emph{individual-based models} juxtaposed with \emph{governing equations}. Individual-based models are not only simple to understand in terms of their fundamental assumptions but have also proven extremely impactful~\cite{Ferguson2020-yn}. However, mathematical tractability is limited, reliable estimates of expectations may require millions of simulations given the fat-tailed, multiplicative nature of epidemics, and inference can be challenging, with parameter inter-dependence making sensitivity analysis unreliable. In contrast, governing equations tend to have a stronger physical interpretation, are easier to perform inference over, and can be embedded in complex models easily \cite{Flaxman2020-lt}. 

The most widely known set of governing equations was presented in the seminal work of Kermack and McKendrick~\cite{Kermack1927}, where they studied the number and distribution of infections of a transmissible disease as it progresses through a population over time. They constructed classes, called \emph{compartments}, and modelled the propagation of infectious disease via interactions among these compartments. The result is the popular \emph{susceptible--infected--recovered} (SIR) model, variants of which are widely used in epidemiology. Stochastic versions of SIR models, formulated either as stochastic differential equations or continuous-time Markov chains, are popular when modelling small populations or stochastic environments \cite{Allen2017-oo}. Deterministic and stochastic SIR models provide an intuitive mechanism for understanding disease transmission, and in the original derivation of~\cite{Kermack1927}, they were noted to be similar to the Volterra equation~\cite{Polyanin1998}. The Volterra equation (of the second kind), or more commonly, the \emph{renewal equation}, is another popular governing equation~\cite{Cauchemez2016,Cori2013,Fraser2007,Nouvellet2018,}. A large body of work in infectious disease epidemiology is based around the renewal equation and many modifications exist~\cite{Aldis2005-co,Champredon2018-sg,Fraser2004-kv,Roberts2004-ec}. There is a connection between specific compartmental models and renewal equations~\cite{Champredon2018,Rizoiu2017} but this link has not been established in full generality. The vast majority of renewal frameworks model only incidence, and the explicit link between prevalence and incidence often requires the use of a latent process for incidence \cite{Brookmeyer1988-ag}. 

Between individual-based and governing equation models are stochastic \emph{branching processes}. Branching processes are applied in the modelling of epidemics by first constructing a stochastic process where infected individuals transmit disease according to simple rules, and then deriving a governing equation for the average behaviour. For example Galton--Watson processes, where individuals infect other individuals at generations specified by a fixed time, provide a tractable and intuitive way of modelling the spread of an infectious disease~\cite{Bartoszynski1967,Getz2006}. In 1948, Bellman and Harris~\cite{Bellman1948} elegantly captured a more complex underlying infection mechanism by formulating an age-dependent branching process, where the age-dependence alludes to individuals who infect other individuals after a random interval of time. Interestingly, the expectation of the Bellman--Harris process \cite{Bellman1948} follows a renewal equation, whereby their framework links the two worlds of individual-based modelling and governing equations. The age-dependence assumption of Bellman and Harris allows, in particular, for the variable time between exposure to a pathogen and subsequent transmission to be modelled more realistically, and provides a framework encoding useful biological characteristics of the infecting pathogen, such as incubation periods and non-monotonic infectiousness. Crump, Mode \cite{Crump1968-rp,Crump1969-gc} and (independently) Jagers \cite{Jagers1975-sd} further extended the Bellman-Harris process to a general branching process where individuals not only can infect at random times, but can do so randomly over the duration of their infection (as opposed to the Bellman-Harris process where all subsequent infections generated by each infected individual happen at a single random time).

\begin{figure}[t]
\includegraphics[width=\textwidth]{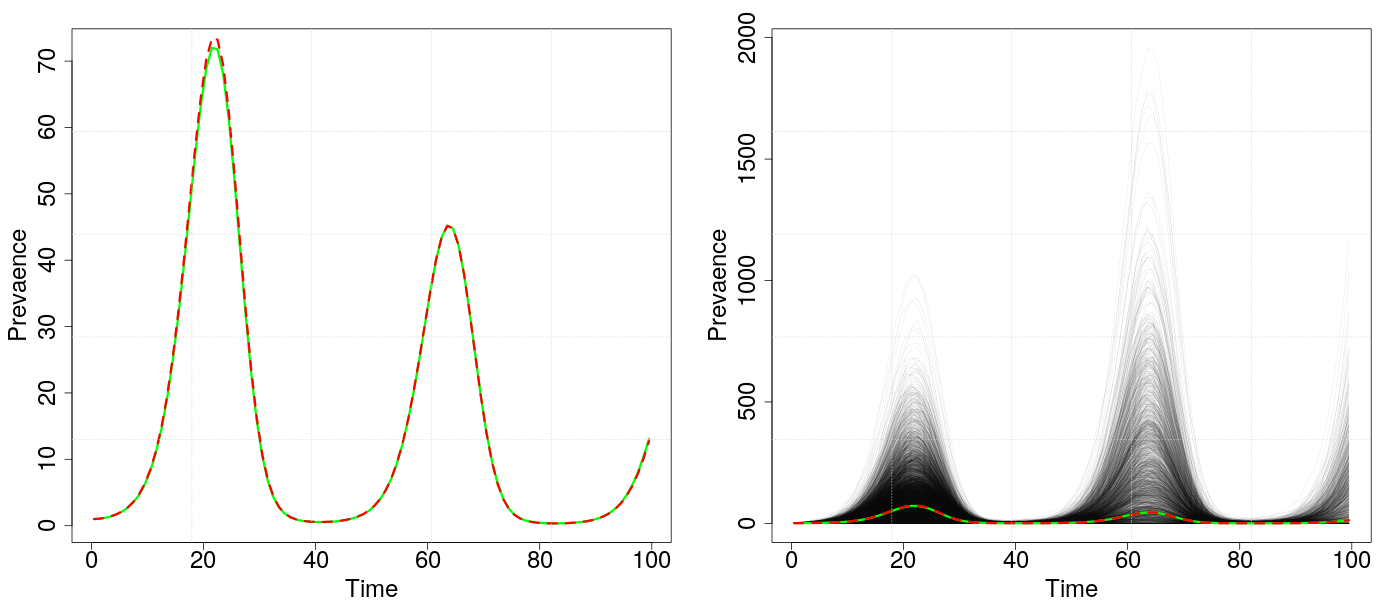}
\centering
\caption{Simulation of an age-dependent Bellman--Harris branching process in terms of prevalence. Left plot shows the Monte Carlo mean (red) alongside the theoretical mean (green). Right plot shows both the Monte Carlo and theoretical mean, overlaid on the underlying 1,000 simulated trajectories (translucent black lines). In this example, the time-varying reproduction number is given by $R(t) = 1.15 + \text{sin}(0.15\,t)$, while the generation interval follows the $ \text{Gamma}(3,1)$ distribution. Algorithm \ref{alg:vector}, given below, is used to compute the theoretical mean.}
\label{fig:Figure1}
\end{figure}

The original formulation by Bellman and Harris~\cite{Bellman1952}, along with subsequent work by Harris~\cite{Harris1963-sa}, the work of Crump, Mode and Jagers~\cite{Crump1968-rp,Crump1969-gc,Jagers1975-sd} as well as the perspective of Bharucha-Reid~\cite{bharucha-reid1956}, with specific application to epidemics, all focused on the simple case of a constant/basic \emph{reproduction number} $R_0$. The form of this renewal equation when only considering $R_0$ is exactly what is commonly used in epidemic modelling where the incidence of infections $\mathrm{I}(t)$ follows a renewal equation given by
\begin{equation*}\label{int:inc-ren}
\mathrm{I}(t) = R_0\int_0^\infty \mathrm{I}(t-u)g(u) \ud u,
\end{equation*}
where $g(\cdot)$ is the probability density function (PDF) of the generation interval.
Introducing a \emph{time-varying} reproduction number $R(t)$ within the Bellman--Harris process in general does not simply entail replacing $R_0$ with $R(t)$ in the renewal equation. This is not possible because a history of how many secondary infections are created is needed. While justifications based on heuristic arguments such as Lotka's~\cite{Lotka1907} (used in tracking the numbers of females in an age-structured population) or the one given by Fraser~\cite{Fraser2007} are valid within their respective contexts, these arguments lose their validity when considering a stochastic age-dependent branching process with a time-varying reproduction process~\cite{Berah2021-qe,Kimmel1983-jo,Kimmel2002-lo}. Indeed, we will demonstrate that these arguments are only valid for the specific case of incidence, not for prevalence or cumulative incidence. Furthermore, to our knowledge, no one has previously investigated a fully time-varying reproduction process under the more general Crump--Mode--Jagers framework. Besides the work by Kimmel~\cite{Kimmel1983-jo} on the time-varying Bellman--Harris process, a \emph{generation-dependent} life length distribution within the Bellman--Harris process has been studied by Fildes~\cite{fildes-1972} and a generation-dependent offspring distribution by Fearn~\cite{fearn-1976}. Moreover, Edler~\cite{edler-1978} and later Biggins and G\"otz~\cite{biggins-goetz-1987} have analysed a generation-dependent reproduction process in the Crump--Mode--Jagers setting.

In this paper, we introduce an outbreak model based on a time-varying version of the Crump--Mode--Jagers process, which we formulate using random characteristics~\cite{Kimmel2002-lo}. Notably, Bellman--Harris, Galton--Watson and Markov branching processes are all special cases of this process. In our novel time-varying Crump--Mode--Jagers process, we specifically allow the statistical properties of infections, i.e., ``offspring'', generated by each individual to vary over time. Building on this model, we lay down a general, stochastic process foundation for incidence, cumulative incidence and prevalence, and characterise the renewal-like integral equations they follow. We show that the equations for prevalence and incidence are consistent with the well-known back-calculation relationship~\cite{Brookmeyer1988-ag,Crump2015-sc} used in infectious disease epidemiology. We also show that the common renewal equation used ubiquitously for modelling incidence \cite{Cori2013,Fraser2007} is in fact, under specific conditions, equivalent to the integral equation for incidence in our framework. Additionally, we formulate a novel reproduction process where infections occur randomly over the duration of each individual's infection according to an inhomogeneous Poisson process. The model thus eschews the common assumption that infections happen instantaneously at a random time, as in the Bellman--Harris process, but still admits analytically tractable integral equations for prevalence and incidence. Finally, we introduce an efficient discretisation algorithm for our newly derived integral equations and use this scheme to estimate rates of transmission from serological prevalence of SARS-CoV-2 in the UK and historical incidence data on Influenza, Measles, SARS and Smallpox.

\section{Model and theoretical results}

\subsection{Time-varying Crump--Mode--Jagers outbreak model}

Throughout the paper, we shall work with an infectious disease outbreak model based on the Crump--Mode--Jagers (CMJ) branching process, which we extend to allow transmission dynamics to vary over time. Our formulation is inspired by Vatutin and Zubkov~\cite{Vatutin1985-ct,Vatutin1993-re}, who give an exposition of the corresponding time-invariant CMJ process using \emph{random characteristics}. In our time-varying CMJ outbreak model, the initial infection occurs at non-random time $\tau \geq 0$.\label{int:tau} All subsequent infections are ``progeny'' of this index case, and we shall denote the set of these infected individuals by $\mathcal{I}^*$. We denote the set of all infected individuals (i.e., including the index case) by $\mathcal{I}$.

The index case corresponds to an individual endowed with a collection of random elements indexed by the infection time,
\begin{equation*}
\{L^\tau, \, \chi^\tau (\cdot), \, N^\tau(\cdot)\}_{\tau \geq 0},
\end{equation*}
where, for any $\tau \geq 0$,
\begin{itemize}
\item $L^\tau$ is a (strictly) positive random variable representing the amount of time the individual remains infected,\label{int:Ltau}
\item $\chi^\tau(\cdot)$ is a stochastic process on $[0,\infty)$ which we shall call the \emph{random characteristic} of the individual, and\label{int:chitau} \item $N^\tau(\cdot)$ is a counting process on $[0,\infty)$ keeping track of the new infections, i.e., ``offspring'', generated by the individual.\label{int:Ntau}
\end{itemize}
For completeness, we set $N^\tau(u) \eqdefl 0 \eqdefr \chi^\tau(u)$ for $u<0$. (We will explain the precise roles of $\chi^\tau(\cdot)$ and $N^\tau(\cdot)$ shortly.) The objects $L^\tau$,  $\chi^\tau(\cdot)$, and $N^\tau(\cdot)$ are typically interdependent, as we shall see below, whilst the interdependence of $(L^\tau,\chi^\tau(\cdot), N^\tau(\cdot))$ and $(L^{\tau'},\chi^{\tau'}(\cdot), N^{\tau'}(\cdot))$ for different $\tau$ and $\tau'$ is in fact immaterial and will be glossed over.
We shall moreover endow each individual $i \in \mathcal{I}^*$ with $\{L^\tau_i, \, \chi^\tau_i(\cdot), \, N^\tau_i(\cdot)\}_{\tau \geq 0}$, which is an independent copy of
$\{L^\tau, \, \chi^\tau(\cdot), \, N^\tau(\cdot)\}_{\tau \geq 0}$. (By an independent copy we mean a new random element which is equal in distribution to the original one and independent of it.)

Suppose now that individual $i \in \mathcal{I}$ is infected at (possibly random) time $\tau_i \geq \tau$.\label{int:taui} Intuitively, the infection time $\tau_i$ then ``selects'' $L^{\tau_i}_i$, $\chi^{\tau_i}_i(\cdot)$, and $N^{\tau_i}_i(\cdot)$ from $\{L^\tau_i, \, \chi^\tau_i(\cdot), \, N^\tau_i(\cdot)\}_{\tau \geq 0}$, which the subsequent infection dynamics of this individual will ``follow.'' (Note that the collection $\{L^\tau_i, \, \chi^\tau_i(\cdot), \, N^\tau_i(\cdot)\}_{\tau \geq 0}$ is independent of the infection time $\tau_i$.) More concretely, $N_i^{\tau_i}(u)$  now stands for the number of new infections generated by the individual $i$ up to time $u+\tau_i$.

\begin{exm}[\textbf{Bellman--Harris process}]\label{exm:BH1}
The Bellman--Harris branching model can informally be characterised, in the context of epidemics, by the principle that each individual generates a random number of new infections which occur \emph{simultaneously} at a random time. Once these new infections have occurred, the individual immediately ceases to be infectious. Let $\xi(\cdot)$ be a stochastic process on $[0,\infty)$ with values in $\N \eqdefl \{0,1,\ldots\}$, independent of $\{L^\tau\}_{\tau \geq 0}$, and then 
define
\begin{equation*}
N^\tau(u) \eqdefl \begin{cases}
0, & u < L^\tau, \\
\xi(\tau+L^\tau), & u \geq L^\tau.
\end{cases}
\end{equation*}
This specification gives rise to the time-varying Bellman--Harris branching process studied by Kimmel \cite{Kimmel1983-jo}. When the distributions of $L^\tau$ and $\xi(t)$ do not depend on the time parameters $\tau \geq 0$ and $t\geq0$, we recover the classical Bellman--Harris process \cite{Bellman1948}. 
\end{exm}

\begin{exm}[\textbf{Inhomogeneous Poisson process model}]\label{exm:Poisson}
In contrast to the Bellman--Harris process, we can consider a more realistic epidemiological model where each infected individual generates new infections randomly and \emph{one by one} according to an inhomogeneous Poisson process until they cease to be infectious. This process, with a constant rate of transmission has been previously studied in the context of the generation time \cite{Svensson2007-wa}.
The infinitesimal rate at time $t$ of new infections generated by an individual originally infected at time $\tau \leq t$ is specified as
\begin{equation*}
\rho(t)k(t-\tau),
\end{equation*}
where $\rho(\cdot)$ is a non-negative function that models population-level variation in transmissibility while $k(\cdot)$ is another non-negative function describing how individual-level infectiousness varies over time \cite{Svensson2007-wa}. For example, specifying $k(t)$ to be low or zero for small $t$ can be used to incorporate an incubation period in the model.    
Let $\Phi(\cdot)$ be a unit-rate, homogeneous Poisson process on $[0,\infty)$, independent of $\{L^\tau\}_{\tau \geq 0}$. Then we can define this model explicitly by
\begin{equation*}
N^\tau(u) \eqdefl \begin{cases}
\Phi \big(\int_0^u \rho(v+\tau) k(v) \ud v\big), & u < L^\tau, \\
\Phi \big(\int_0^{L^\tau} \rho(v+\tau) k(v) \ud v\big), & u \geq L^\tau.
\end{cases}
\end{equation*}
(If $\rho(t) \equiv \rho$ and $k(t) \equiv k$, both constant, then new infections follow a homogeneous Poisson process with rate $\rho k$  until the individual is no longer infected.). 
\end{exm}

\begin{exm}[\textbf{L\'evy and Cox process models}]\label{exm:Levy-Cox}
In the inhomogeneous Poisson process model of Example \ref{exm:Poisson}, tractability does not hinge on the assumption that $\Phi(\cdot)$ is a Poisson process. We could in fact replace it with a more general, integer-valued L\'evy process (i.e., a process with independent and identically distributed increments), where jumps need not be of unit size (e.g., a compound Poisson process). Similarly, replacing the deterministic function $\rho(\cdot)$ with a stochastic process, as long as it is independent of $\Phi(\cdot)$, would be straightforward. In the Poisson case, this would turn $N^\tau(\cdot)$ into to a doubly-stochastic Cox process. However, for simplicity and concreteness, we shall stick to the simpler setting of Example \ref{exm:Poisson}.
\end{exm}
\begin{figure}[t]
\includegraphics[scale=0.55]{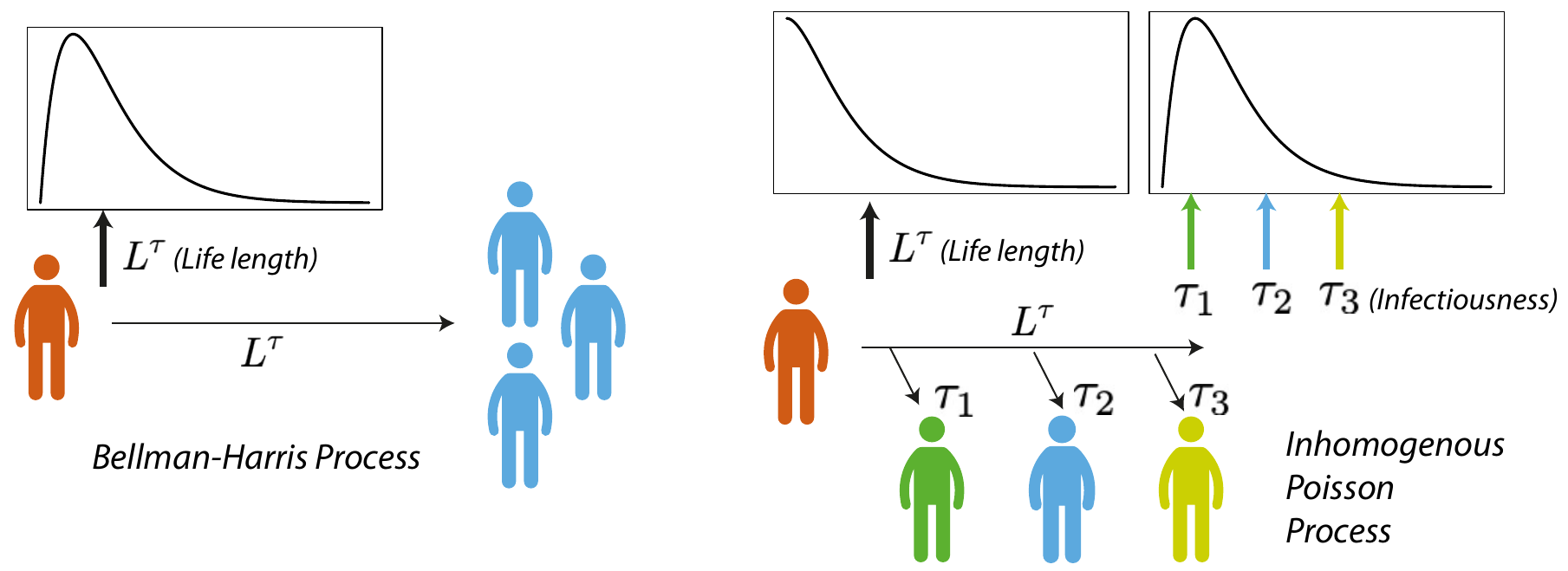}
\centering
\caption{Schematic of infections generated under a Bellman--Harris process and an inhomogeneous Poisson process model. In a Bellman--Harris process, after a generation interval has elapsed, new infections happen at the same time (instantaneously). In the inhomogeneous Poisson process model, an individual is infectious for a period, over which their infectiousness varies, and they produce infections one by one.}
\label{fig:Figure0}
\end{figure}
\begin{rem}[\textbf{Epidemiological interpretation of $L^\tau$ and $k$}]
In the Bellman--Harris process of Example \ref{exm:BH1}, $L^\tau$ is directly interpreted as the generation interval \cite{Svensson2007-wa}, that is, the time taken for the secondary cases to be infected by a primary case. In the Bellman--Harris process all infections happen at the same time --- for example in Figure \ref{fig:Figure0} we have $\xi(\tau + L^\tau)=3$, after $L^\tau$ time units has elapsed since the index case was infected at time $\tau$. In contrast, in the inhomogenous Poisson process model of Example \ref{exm:Poisson} (and also the L\'evy and Cox process models of Example \ref{exm:Levy-Cox}), $L^\tau$ corresponds to how long an individual remains infected (the duration of infection). During this period, an individual can infect others with rate that depends on $\rho(\cdot)$, which describes the calendar-time variation of overall infectiousness in the population, and on $k(\cdot)$, which in turn describes how the infectiousness of each infected individual varies over the course of their infection. The individual's infectiousness profile $k(\cdot)$ can be set as constant, i.e., variation in the individual's infectiousness is only due to calendar-time variation in overall infectiousness. If $k(\cdot)$ is specified to vary significantly, by contrast, then it is advisable to ensure that infections are most likely to end when infectiousness is low. Concretely, this means that $k(\cdot)$ should then be paired with $L^\tau$ such that the bulk of its distribution coincides with low values of $k(\cdot)$, as in the empirical application in Section \ref{ssec:Empirical} below.
\end{rem}

The random characteristic is used merely as a book-keeping device, to keep track of an individual's infection status in two ways --- whether they have been infected in the past or, alternatively, whether they are infected at the moment. It is fundamental to obtaining a unified derivation of both cumulative incidence and prevalence in what follows.
\begin{exm}[\textbf{Cumulative incidence and prevalence}]\label{exm:char}
The random characteristic (in fact non-random!)
\begin{equation}\label{eq:cinc}
\chi^\tau(u) \eqdefl \begin{cases}
0, & u < 0,\\
1, & u \geq 0,
\end{cases}
\end{equation}
determines whether the individual has been infected by time $u+\tau$ and is therefore used to derive cumulative incidence. The random characteristic
\begin{equation}\label{eq:prev}
\chi^\tau(u) \eqdefl \begin{cases}
0, & u < 0,\\
1, & u \in [0,L^\tau),\\
0, & u > L^\tau,
\end{cases}
\end{equation}
determines whether the individual remains infected at time $u+\tau$ and is  used to derive prevalence.
\end{exm}

\subsection{Cumulative incidence and prevalence}

We will now derive integral equations for cumulative incidence and prevalence under this model.
To this end, we study the stochastic process\label{int:Zproc}
\begin{equation*}
Z(t,\tau) \eqdefl \sum_{i \in \mathcal{I}} \chi^{\tau_i}_i(t-\tau_i),\quad t \geq \tau \geq 0,
\end{equation*}
recalling that $\tau$ is the infection time of the index case. 
For the given random characteristic \eqref{eq:cinc}, $Z(t,\tau)$ counts the number of infections occurred by time $t$ and for \eqref{eq:prev} the number of infected individuals at time $t$, respectively. Our goal is to derive an equation for the expectation of $Z(t,\tau)$, covering both cases.

Before embarking on the derivation of the equation governing $\E[Z(t,\tau)]$, we shall first introduce technical assumptions ensuring $\E[Z(t,\tau)]<\infty$, which a fortiori guarantees that $Z(t,\tau)$ is finite with probability one, a property known as \emph{regularity} in the branching process literature \cite{sevastjanov-1967}. Regarding $N^\tau(\cdot)$, we write
\begin{equation*}
\Lambda^{\tau}(u)  \eqdefl \E[N^\tau(u)], \quad \tau \geq 0, \quad u \geq 0,
\end{equation*}
and \label{int:Lambda} henceforth assume that there is a non-decreasing, right-continuous function $\overline{\Lambda} : [0,\infty) \rightarrow [0,\infty)$ such that
\begin{equation}\label{eq:regularity}
\overline{\Lambda}(0) < 1 \quad \text{and} \quad \Lambda^\tau(u) \leq \overline{\Lambda}(u) \quad \text{for any $\tau \geq 0$ and $u \geq 0$.}
\end{equation}
(We will give sufficient conditions that imply this assumption in the context of Examples \ref{exm:BH1} and \ref{exm:Poisson} below in Examples \ref{exm:BH2} and \ref{exm:Poisson-2}, respectively.) Moreover, we assume that the random characteristic $\chi^\tau(\cdot)$ satisfies $0 \leq \chi^\tau(u) \leq 1$ for any $\tau \geq 0$ and $u \geq 0$, which evidently accommodates both \eqref{eq:cinc} and \eqref{eq:prev} from Example \ref{exm:char}.
Under these assumptions, straightforward adaptation of the proof of Lemma 4.2 in \cite{Crump1968-rp} (cf.\ the proof of Theorem 2.1 in \cite{edler-1978}) yields $\E[Z(t,\tau)]<\infty$ for any $t \geq \tau \geq 0$.

Now, singling out the index case, we can write
\begin{equation}\label{eq:first-gen-1}
Z(t,\tau) = \chi^\tau(t-\tau) + \sum_{i \in \mathcal{I}^*} \chi^{\tau_i}_i(t-\tau_i).
\end{equation}
The key insight in the analysis of \eqref{eq:first-gen-1} is to stratify the infected individuals in $\mathcal{I}^*$ according to their (unique) ``ancestor'' among the individuals infected by the index case. More concretely, let $i_1,i_2,\ldots \in \mathcal{I}^*$ label  the ``offspring'' of the index case in chronological order, i.e., so that $\tau \leq \tau_{i_1} \leq \tau_{i_2} \leq \cdots$, and let $\mathcal{I}_{k} \subset \mathcal{I}^*$ for each $k = 1,2,\ldots$ denote the set consisting of $i_k$ and its ``progeny.'' We can then write 
\begin{equation*}
\sum_{i \in \mathcal{I}^*} \chi^{\tau_i}_i(t-\tau_i) = \sum_{k \,:\, \tau_{i_k} \leq t} \underbrace{\sum_{i \in \mathcal{I}_k} \chi^{\tau_i}_i(t-\tau_i)}_{\eqdefr Z_k(t)}.
\end{equation*}
This is an analogue of the \emph{principle of first generation} for the Bellman--Harris process \cite[Theorem 6.1]{Harris1963-sa} (see also \cite[p.~5]{Kimmel1983-jo}).

Conditional on the random times $\tau_{i_1},\tau_{i_2},\ldots$, the random variables $Z_1(t),Z_2(t),\ldots$ can be shown to be mutually independent, with $Z_k(t)$ equal in distribution to $\widetilde{Z}(t,\tau_{i_k})$, where $\big\{\widetilde{Z}(\cdot,\tau)\big\}_{\tau \geq 0}$ is an independent copy of $\{Z(\cdot,\tau)\}_{\tau \geq 0}$ (independent of $\tau_{i_1},\tau_{i_2},\ldots$, in particular). Thus,\label{f-int}
\begin{equation*}
f(t,\tau) \eqdefl \E[Z(t,\tau)] = \E[\chi^\tau(t-\tau)] + \E\Bigg[ \sum_{k \,:\, \tau_{i_k} \leq t} Z_k(t)\Bigg],
\end{equation*}
where, using the law of total expectation,
\begin{equation*}
\begin{split}
\E\Bigg[ \sum_{k \,:\, \tau_{i_k} \leq t} Z_k(t)\Bigg] & = \E\Bigg[\E\Bigg[ \sum_{k \,:\, \tau_{i_k} \leq t} Z_k(t)\,\Bigg| \,\tau_{i_1},\tau_{i_2},\ldots \Bigg]\Bigg] \\
& = \E\Bigg[ \sum_{k \,:\, \tau_{i_k} \leq t} \E[ Z_k(t)\,| \,\tau_{i_1},\tau_{i_2},\ldots ]\Bigg] \\
& = \E\Bigg[ \sum_{k \,:\, \tau_{i_k} \leq t} \E\big[\widetilde{Z}(t,\tau)\big]_{\tau = \tau_{i_k}}\Bigg].
\end{split}
\end{equation*}
Since $\big\{\widetilde{Z}(t,\tau)\big\}_{\tau \geq 0}$ is equal in distribution to $\{Z(t,\tau)\}_{\tau \geq 0}$, we get
\begin{equation*}
\begin{split}
\E\Bigg[ \sum_{k \,:\, \tau_{i_k} \leq t} \E\big[\widetilde{Z}(t,\tau)\big]_{\tau = \tau_{i_k}}\Bigg] & = \E\Bigg[ \sum_{k \,:\, \tau_{i_k} \leq t} f(t,\tau_{i_k})\Bigg] \\
& = \E\Bigg[ \sum_{v \in (\tau,t]} f(t,v) \Delta N^\tau(v-\tau) \Bigg] \\
& = \E\Bigg[ \sum_{u \in (0,t-\tau]} f(t,u+ \tau) \Delta N^\tau(u) \Bigg] \\
& = \E\Bigg[ \int_{(0,t-\tau]} f(t,u+\tau) \ud N^\tau(u) \bigg] \\
& = \int_{(0,t-\tau]} f(t,u+ \tau) \E[\ud N^\tau(u)] \\
& = \int_{(0,t-\tau]} f(t,u+\tau) \ud \Lambda^\tau(u),
\end{split}
\end{equation*}
where 
$\Delta N^\tau(u)  \eqdefl N^\tau(u)-\lim_{v \rightarrow u-}N^\tau(v)$ denotes the jump size of $N^\tau(\cdot)$ at time $u \geq 0$. Therefore, the function $(t,\tau) \mapsto f(t,\tau)$ is governed by the integral equation
\begin{equation}\label{eq:integral-eq}
f(t,\tau) = \E[\chi^\tau(t-\tau)] + \int_{(0,t-\tau]} f(t,u+\tau) \ud \Lambda^\tau (u), \quad t \geq \tau \geq 0.
\end{equation}

For the random characteristic \eqref{eq:cinc}, $f(t,\tau)$ is the cumulative incidence \label{int:cinc} at time $t$, and we shall denote it by $\mathrm{CI}(t,\tau)$. Since $\E[\chi^\tau(t-\tau)] = 1$ in this case for $t \geq \tau$, the equation \eqref{eq:integral-eq} transforms into
\begin{equation}\label{eq:cinc-eq}
\mathrm{CI}(t,\tau) = 1 + \int_{(0,t-\tau]} \mathrm{CI}(t,u+\tau)  \ud \Lambda^\tau (u).
\end{equation}
In the case \eqref{eq:prev}, $f(t,\tau)$ is the prevalence \label{int:prev} at time $t$, which we henceforth denote by $\mathrm{Pr}(t,\tau)$. In this case,
\begin{equation*}
\E[\chi^\tau(t-\tau)] = \prob[t-\tau < L^\tau] = 1 - G^\tau(t-\tau),\quad t \geq \tau,
\end{equation*}
where $G^\tau(\cdot)$ \label{int:Gtau} denotes the cumulative distribution function (CDF) of $L^\tau$. Writing $\overline{G}^\tau(\cdot) \eqdefl 1 - G^\tau(\cdot)$ \label{int:Gtaubar} for the survival function associated with $G^\tau(\cdot)$, we have then
\begin{equation}\label{eq:prev-eq}
\mathrm{Pr}(t,\tau) = \overline{G}^\tau(t-\tau) + \int_{(0,t-\tau]} \mathrm{Pr}(t,u+\tau) \ud \Lambda^\tau (u).
\end{equation}

\begin{exm}[\textbf{Bellman--Harris process, cont'd}]\label{exm:BH2} Let us consider the Bellman--Harris case of Example \ref{exm:BH1} 
 and write $R(t) \eqdefl \E[\xi(t)]$ for the (time-varying) \emph{reproduction number} at time $t \geq 0$. Let us also denote the indicator function of a set $A$ by $\mathbf{1}_A$.\label{int:ind} Using the law of total expectation and the independence between $\xi(\cdot)$ and $\{ L^\tau \}_{\tau \geq 0}$, we get then
\begin{align}
\Lambda^\tau(u) = \E[N^\tau(u)] & = \E[\xi(L^\tau+\tau)\mathbf{1}_{\{ u \geq L^\tau \}}] \notag \\ 
&= \E[ \E[\xi(L^\tau+\tau)\, | \, L^\tau]\mathbf{1}_{\{ u \geq L^\tau \}}] \notag \\
&= \int_{(0,u]} R(u'+\tau) \ud G^\tau(u').\label{eq:BH-Lambda}
\end{align}
We will henceforth assume that the maximal reproduction number $\overline{R} \eqdefl \sup_{t \geq 0} R(t)$ is finite and that the function $\widehat{G}(u) \eqdefl \lim_{v \rightarrow u+} \sup_{\tau \geq 0} G^\tau(v)$, $u \geq 0$, satisfies
\begin{equation}\label{eq:Rmax}
\widehat{G}(0) < \overline{R}^{-1}.
\end{equation}
Intuitively, this condition ensures that the distribution of $L^\tau$ does not become too concentrated near zero over time.
We can then define a non-decreasing, right-continuous function $\overline{\Lambda}(u) \eqdefl \overline{R} \widehat{G}(u)$, $u \geq 0$ which, in view of \eqref{eq:BH-Lambda} and \eqref{eq:Rmax}, satisfies the assumption \eqref{eq:regularity}. Hence, we deduce $\E[Z(t,\tau)]<\infty$, and the regularity of the branching process follows.
Inserting \eqref{eq:BH-Lambda} into \eqref{eq:cinc-eq} and \eqref{eq:prev-eq}, respectively, we obtain
\begin{align*}
\mathrm{CI}(t,\tau) &= 1 + \int_{(0,t-\tau]} \mathrm{CI}(t,u+\tau) R(u+\tau) \ud G^\tau(u),\\
\mathrm{Pr}(t,\tau) &= \overline{G}^\tau(t-\tau) + \int_{(0,t-\tau]} \mathrm{Pr}(t,u+\tau) R(u+\tau) \ud G^\tau(u),
\end{align*}
which agree with \cite[Theorem 5.1]{Kimmel1983-jo}. When $G^\tau(\cdot)$ admits a PDF $g^\tau(\cdot)$,\label{int:gtau} the most relevant case in practice, we can simplify the equations further to
\begin{align*}
\mathrm{CI}(t,\tau) &= 1 + \int_0^{t-\tau} \mathrm{CI}(t,u+\tau) R(u+\tau)  g^\tau(u) \ud u,\\
\mathrm{Pr}(t,\tau) &= \overline{G}^\tau(t-\tau) + \int_0^{t-\tau} \mathrm{Pr}(t,u+\tau) R(u+\tau) g^\tau(u) \ud u.
\end{align*}
\end{exm}

\begin{exm}[\textbf{Inhomogeneous Poisson process model, cont'd}]\label{exm:Poisson-2}
To analyse the Poisson process model of Example \ref{exm:Poisson}, we note first that for any $u \geq 0$, 
\begin{equation*}
\begin{split}
\Lambda^\tau(u) = \E[N^\tau(u)] & = \E\bigg[\Phi \bigg(\int_0^u \rho(v+\tau)k(v) \ud v\bigg)\mathbf{1}_{\{ u < L^\tau \}}\bigg] \\ & \quad + \E\bigg[\Phi \bigg(\int_0^{L^\tau} \rho(v+\tau)k(v) \ud v\bigg)\mathbf{1}_{\{ u \geq L^\tau \}}\bigg],
\end{split}
\end{equation*}
whence 
\begin{equation*}
\Lambda^\tau(u) \leq \E\bigg[\Phi \bigg(\int_0^u \rho(v+\tau)k(v) \ud v\bigg)\bigg] \leq \int_0^u \sup_{\tau \geq 0}\rho(v+\tau)k(v) \ud v \eqdefr \overline{\Lambda}(u).
\end{equation*}
If we assume, say, that $\rho(\cdot)$ and $k(\cdot)$ are bounded, then $\overline{\Lambda}(\cdot)$ is non-decreasing, continuous and satisfies \eqref{eq:regularity}, implying $\E[Z(t,\tau)]<\infty$ and the regularity of the branching process. (The CDF $G^\tau(\cdot)$ does not play a role in regularity for this model since, unlike in the Bellman--Harris process, the random variable $L^\tau$ cannot precipitate secondary infections.)
To work out an expression for $\Lambda^\tau(u)$, we shall further assume that $G^\tau(\cdot)$ admits a PDF $g^\tau(\cdot)$, as above.
Invoking the independence between $\Phi(\cdot)$ and $\{ L^\tau \}_{\tau \geq 0}$ and the law of total expectation, we obtain
\begin{align}
\E\bigg[\Phi \bigg(\int_0^u \rho(v+\tau)k(v) \ud v\bigg)\mathbf{1}_{\{ u < L^\tau \}}\bigg] & = \E\bigg[\Phi \bigg(\int_0^u \rho(v+\tau)k(v) \ud v\bigg)\bigg] \prob[u < L^\tau] \notag \\
& = \overline{G}^\tau (u) \int_0^u \rho(v+\tau)k(v) \ud v \label{eq:Poisson-2}
\end{align}
and
\begin{equation*}
\begin{split}
\E\bigg[\Phi \bigg(\int_0^{L^\tau} \rho(v+\tau)k(v) \ud v\bigg)\mathbf{1}_{\{ u \geq L^\tau \}}\bigg] & = \E\bigg[\E\bigg[\Phi \bigg(\int_0^{L^\tau} \rho(v+\tau)k(v) \ud v\bigg)\mathbf{1}_{\{ u \geq L^\tau \}}\bigg| \, L^\tau\,\bigg]\bigg] \\
& = \E\bigg[\E\bigg[\Phi \bigg(\int_0^{\ell} \rho(v+\tau)k(v) \ud v\bigg)\bigg]_{\ell = L^\tau}\mathbf{1}_{\{ u \geq L^\tau \}} \bigg] \\
& = \int_0^u \int_0^{u'} \rho(v+\tau)k(v) \ud v \, g^\tau (u') \ud u'.
\end{split}
\end{equation*}
Integrating \eqref{eq:Poisson-2} by parts,
\begin{equation*}
\begin{split}
\overline{G}^\tau (u) \int_0^u \rho(v+\tau)k(v) \ud v & = \int_0^u \int_0^{u'} \rho(v+\tau)k(v) \ud v \, \ud \overline{G}^\tau (u') + \int_0^u  \overline{G}^\tau (u') \rho(u'+\tau)k(u')  \ud u' \\
& = -\int_0^u \int_0^{u'} \rho(v+\tau)k(v) \ud v \, g^\tau (u') \ud u' + \int_0^u   \rho(u'+\tau)k(u') \overline{G}^\tau (u')  \ud u',
\end{split}
\end{equation*}
since $\frac{\ud \overline{G}^\tau (u)}{\ud u} = -\frac{\ud G^\tau (u)}{\ud u} = -g^\tau (u)$. Therefore,
\begin{equation*}
\Lambda^\tau(u) = \int_0^u   \rho(u'+\tau)k(u') \overline{G}^\tau (u')  \ud u',
\end{equation*}
whereby the equations for cumulative incidence and prevalence read as
\begin{align}
\mathrm{CI}(t,\tau) &= 1 + \int_0^{t-\tau} \mathrm{CI}(t,u+\tau) \rho(u+\tau) k(u) \overline{G}^\tau(u) \ud u,\label{eq:CI-Poisson}\\
\mathrm{Pr}(t,\tau) &= \overline{G}^\tau(t-\tau) + \int_0^{t-\tau} \mathrm{Pr}(t,u+\tau) \rho(u+\tau)k(u) \overline{G}^\tau(u) \ud u,\label{eq:Pr-Poisson}
\end{align}
respectively, in this case.
\end{exm}

\begin{exm}[\textbf{Alternative Poisson process model}]
An alternative version of the inhomogeneous Poisson process model, suggested by an anonymous referee, can be formulated by assuming that an infected individual's infectiousness evolves at a time scale determined by $L^\tau$ via
\begin{equation*}
\tilde{k}\bigg(\frac{u}{L^\tau}\bigg), \quad u \geq 0,   
\end{equation*}
where $\tilde{k}(\cdot)$ is a continuous function, (strictly) positive on the interval $[0,1]$ and zero elsewhere. Concretely, we then define
\begin{equation*}
N^\tau(u) \eqdefl \Phi\bigg( \int_0^u \rho(v+\tau) \tilde{k}\bigg(\frac{v}{L^\tau}\bigg) \ud v \bigg), \quad u \geq 0,     
\end{equation*}
where $\rho(\cdot)$ and $\Phi(\cdot)$ are as in Examples \ref{exm:Poisson} and \ref{exm:Poisson-2}. Note that, by the properties of $\tilde{k}(\cdot)$, we have $N^\tau(u) = N^\tau(L^\tau)$ for $u \geq L^\tau$. A straightforward computation shows that
\begin{equation*}
\Lambda^\tau(u) = \int_0^u \rho(u'+\tau) \tilde{g}^\tau(u') \ud u',    
\end{equation*}
where
\begin{equation*}
    \tilde{g}^\tau(u) \eqdefl \int_0^\infty \tilde{k}\bigg(\frac{u}{v}\bigg) g^\tau(v) \ud v, \quad u \geq 0.
\end{equation*}
Since $\tilde{k}(\cdot)$ is necessarily bounded, the regularity of the resulting branching process is guaranteed. Moreover, we note that cumulative incidence and prevalence for the model can be analysed along the lines of the original Poisson process model simply by substituting $k(u) \overline{G}^\tau(u)$ with $\tilde{g}^\tau(u)$ in the integral equations \eqref{eq:CI-Poisson} and \eqref{eq:Pr-Poisson}.
\end{exm}

\begin{rem}[\textbf{Probability generating functions}]
In the Bellman--Harris case of Examples \ref{exm:BH1} and \ref{exm:BH2}, we can also analyse the distribution of $Z(t,\tau)$ via its \emph{generating function} $\phi(s;t,\tau) \eqdefl \E[s^{Z(t,\tau)}]$, $s \in [-1,1]$, letting us study, e.g., higher moments. Concretely, one can show that $\phi(\,\cdot\,;t,\tau)$ satisfies the integral equations
\begin{align*}
\phi(s;t,\tau) & = s \,\overline{G}^\tau(t-\tau) + s\int_{(0,t-\tau]} \psi\big(\phi(s;t,u+\tau);u+\tau\big) \ud G^\tau(u), \\
\phi(s;t,\tau) & = s \,\overline{G}^\tau(t-\tau) + \int_{(0,t-\tau]} \psi\big(\phi(s;t,u+\tau);u+\tau\big) \ud G^\tau(u),
\end{align*}
for random characteristics \eqref{eq:cinc} and \eqref{eq:prev}, respectively, where $\psi(s;t) \eqdefl \E[s^{\xi(t)}]$, $s \in [-1,1]$. These are special cases of \cite[Equations (3.3) and (3.4)]{Kimmel1983-jo}, whilst self-contained re-derivations in the case where $G^\tau(\cdot)$ does not depend on the infection time $\tau$ are given in \cite{Bellman1948}.
\end{rem}
\begin{rem}[\textbf{Relationship between $\rho(t)$ and $R(t)$}]\label{rem:case_reproduction_numer} The quantity
$R(t)$ in the context of the Bellman--Harris process (Examples \ref{exm:BH1} and \ref{exm:BH2}) is more precisely the \emph{instantaneous} reproduction number, i.e., the expected number of secondary cases arising from a primary case when those infections occur at time $t$. In the context of a real-time epidemic, $R(t)$ is generally interpreted as the average number of secondary cases that would arise from a primary case infected at time $t$ if conditions remained the same after time $t$ \cite{Fraser2007}. The quantity $\rho(t)$ in the Poisson process model (Examples \ref{exm:Poisson} and \ref{exm:Poisson-2}), in contrast, is a time varying transmission rate, i.e., scaled by time, and therefore exists on a different scale. An alternative way of analysing $R(\cdot)$ is to use the \emph{case reproduction number} $\mathcal{R}(t)$ \cite{Gostic2020-mm,Wallinga2004}, which represents the average number of secondary cases arising from a primary case infected at time $t$, i.e., transmissibility after time $t$. It is similarly possible to also analyse $\rho(\cdot)$ through the case reproduction number and therefore compare the rates of transmission in both models commensurably. Namely, given $\rho(\cdot)$ and $R(\cdot)$, they can be transformed into $\mathcal{R}(\cdot)$ and be comparable on the same scale via 
\begin{align*}
    \mathcal{R}_{\mathrm{Pois}}(t)&=\int_t^\infty \rho(u)k(u-t)\overline{G}^t(u-t) \ud u,\\
    \mathcal{R}_{\mathrm{BH}}(t)&=\int_t^\infty R(u)g^t(u-t) \ud u.
\end{align*}

\end{rem}

\begin{rem}[\textbf{When do the Bellman--Harris process and the Poisson process model agree?}]\label{rem:epi_equivalence}
The fundamental difference between the Bellman--Harris (Example \ref{exm:BH2}) and the Poisson process model (Example \ref{exm:Poisson-2}) integral equations is that the Bellman--Harris integral equations are parameterised by $g^\tau(\cdot)$, and the Poisson process model equations by $k(\cdot)\overline{G}^\tau(\cdot)$. Within the Bellman--Harris process, the precise interpretation of $g^\tau(\cdot)$ is the PDF of the time between an individual becoming infected and occurrence of \emph{all} subsequent infections generated by the individual, i.e., the generation time or interval \cite{Svensson2007-wa}. In contrast, the Poisson process model is parameterised by the product of the infectiousness profile $k(\cdot)$, which broadly corresponds to the generation time \cite{Cori2012-ji}, and the survival function $\overline{G}^\tau(\cdot)$ of the duration of the infection. Generally, these two models differ in terms of their behaviour. That said, they give rise to equivalent cumulative incidence and prevalence provided
\begin{equation}\label{eq:model-equiv}
k(u) = \frac{g^\tau(u)}{\overline{G}^\tau(u)}, \quad u \geq 0, \quad \tau \geq 0.
\end{equation}
Hence, cumulative incidence and prevalence roughly agree between the two models when the infectiousness profile $k(\cdot)$ approximates the \emph{hazard function} of $L^\tau$, i.e., the right-hand side of \eqref{eq:model-equiv}. Even in this case, the higher moments of the models typically do not agree, however.
\end{rem}

\subsection{Incidence}

Incidence is defined as the time-derivative of cumulative incidence. To derive an integral equation for incidence \`a la \eqref{eq:cinc-eq} and \eqref{eq:prev-eq}, we shall assume that the function $\Lambda^\tau(\cdot)$ is continuously differentiable, that is,
\begin{equation}\label{eq:lambda-dens}
\Lambda^\tau(u) = \int_0^u \lambda^\tau(u') \ud u',
\end{equation}
for some continuous function $\lambda^\tau(\cdot)$\label{int:lambda}. The function $\lambda^\tau(\cdot)$ is necessarily non-negative since $N^\tau(\cdot)$ is a counting process. The assumption \eqref{eq:lambda-dens} rules out infections occurring in a discrete time grid. It is satisfied with $\lambda^\tau(u) = \rho(u+\tau)k(u) \overline{G}^\tau(u)$ in Example \ref{exm:Poisson-2} provided $\rho(\cdot)$ and $k(\cdot)$ are continuous, and with $\lambda^\tau(u) = R(u+\tau) g^\tau(u)$ in Example \ref{exm:BH2} provided $R(\cdot)$ is continuous and $G^\tau(\cdot)$ has a continuous PDF $g^\tau(\cdot)$.

Cumulative incidence, by definition, equals zero before the  index case is infected at time $\tau$, whilst it then jumps to one. Hence, cumulative incidence, when understood as a function on the entire real line, satisfies
\begin{equation}\label{eq:CI-R}
\mathrm{CI}(t,\tau) = \mathbf{1}_{[0,\infty)}(t-\tau) + \int_0^{t-\tau} \mathrm{CI}(t,u+\tau) \lambda^\tau(u) \ud u, \quad t \in \R.
\end{equation}
(When $t < \tau$ we will interpret the integral, and similar integrals in what follows, as zero.) Incidence \label{int:inc} is then defined as the time-derivative
\begin{equation*}
\mathrm{I}(t,\tau) \eqdefl \frac{\partial}{\partial t} \mathrm{CI}(t,\tau).
\end{equation*}
Before deriving incidence in full generality, let us however study the time-derivative of a related quantity
\begin{equation*}
\widetilde{\mathrm{CI}}(t,\tau) \eqdefl \mathrm{CI}(t,\tau) - \mathbf{1}_{[0,\infty)}(t-\tau), \quad t \in \R,    
\end{equation*}
which omits the initial jump and, in view of \eqref{eq:CI-R}, satisfies
\begin{equation}\label{eq:CI-tilde}
\widetilde{\mathrm{CI}}(t,\tau) =  \int_0^{t-\tau} \lambda^\tau (u) \ud u + \int_0^{t-\tau} \widetilde{\mathrm{CI}}(t,u+\tau) \lambda^\tau(u) \ud u.
\end{equation}
Applying the Leibniz integral rule to the second integral on the right-hand side of \eqref{eq:CI-tilde} formally (see Remark \ref{rem:rigorous} below), we obtain
\begin{equation}\label{eq:Leibniz}
\frac{\partial}{\partial t} \widetilde{\mathrm{CI}}(t,\tau) = \lambda^\tau(t-\tau) + \int_0^{t-\tau} \frac{\partial}{\partial t} \widetilde{\mathrm{CI}}(t,u+\tau) \lambda^\tau (u) \ud u - \underbrace{\widetilde{\mathrm{CI}}(t,t)}_{=0} \lambda^\tau(t-\tau).  
\end{equation}
Since $\mathrm{I}(t,\tau) = \frac{\partial}{\partial t} \mathrm{CI}(t,\tau) = \frac{\partial}{\partial t} \widetilde{\mathrm{CI}}(t,\tau)$ for $t > \tau$, we deduce that
\begin{equation}\label{eq:inc-eq}
\mathrm{I}(t,\tau) = \lambda^\tau(t-\tau) + \int_{(0,t-\tau)} \mathrm{I}(t,u+\tau) \lambda^\tau (u) \ud u, \quad t > \tau.    
\end{equation}
We have taken $(0,t-\tau)$ as the integration domain since $\frac{\partial}{\partial t} \widetilde{\mathrm{CI}}(t,u+\tau)$ and $\mathrm{I}(t,u+\tau)$ do not agree at $u=t-\tau$ for reasons that will become clear in the next paragraph.

Whilst \eqref{eq:inc-eq} already describes incidence for $t > \tau$, for further developments in Sections \ref{sec:backcalculation} and \ref{sec:consistency} it is essential that we have an equation characterising incidence for any $t \geq \tau$. Thus, we need to also deal with the case $t = \tau$ where the time-derivative $\frac{\partial}{\partial t} \mathrm{CI}(t,\tau)$ cannot be defined in the classical sense due to the jump in cumulative incidence.
To this end, it is helpful to note that the derivative of $t \mapsto \mathbf{1}_{[0,\infty)}(t-\tau)$ may be understood as a Dirac delta function $\delta(\,\cdot\, - \tau)$ \label{int:delta} in a distributional sense. We recall that the Dirac delta function is a generalised function with the characteristic property $\int_{\R} f(x) \delta(y-x) \ud x = f(y)$. Now,
\begin{equation*}
\mathrm{I}(t,\tau) = \delta(t-\tau) + \frac{\partial}{\partial t} \widetilde{\mathrm{CI}}(t,\tau), \quad t \in \R.    
\end{equation*}
In particular, formally
\begin{equation}\label{eq:inc-zero}
\mathrm{I}(\tau,\tau) = \delta(0) + \lambda^\tau(0).    
\end{equation}
Note that
\begin{equation*}
\begin{split}
\lambda^\tau(t-\tau) & = \int_{\{t-\tau\}} \delta\big(t-(u+\tau)\big) \lambda^\tau(u) \ud u \\
& = \int_{\{t-\tau\}} \bigg(\delta\big(t-(u+\tau)\big) + \frac{\partial}{\partial t} \widetilde{\mathrm{CI}}(t,u+\tau)\bigg) \lambda^\tau(u) \ud u \\
& = \int_{\{t-\tau\}} \mathrm{I}(t,u+\tau) \lambda^\tau(u) \ud u,
\end{split}
\end{equation*}
since $\int_{\{t-\tau\}} \frac{\partial}{\partial t} \widetilde{\mathrm{CI}}(t,u+\tau) \lambda^\tau(u) \ud u = 0$. Thus, we can write the right-hand side of \eqref{eq:inc-eq} as a single integral over $(0,t-\tau]$, i.e.,
\begin{equation*}
\mathrm{I}(t,\tau) = \int_{(0,t-\tau]} \mathrm{I}(t,u+\tau) \lambda^\tau (u) \ud u, \quad t > \tau.    
\end{equation*}
Consequently, we find that incidence is generally governed by the equation
\begin{equation}\label{eq:inc-eq-2}
\mathrm{I}(t,\tau) = \delta(t-\tau) + \int_{[0,t-\tau]} \mathrm{I}(t,u+\tau) \lambda^\tau (u) \ud u, \quad t \geq \tau.   
\end{equation}
\begin{rem}
In \eqref{eq:inc-eq-2}, we have adjusted the integration domain from $(0,t-\tau]$ to $[0,t-\tau]$ to ensure that the equation agrees with \eqref{eq:inc-zero} for $t = \tau$. (This adjustment is immaterial for $t > \tau$.) To see why this is the case, note that the right-hand side of \eqref{eq:inc-eq-2} consists of the generalised function $\delta(t-\tau)$ and the integral $\int_{[0,t-\tau]} \mathrm{I}(t,u+\tau) \lambda^\tau (u) \ud u$, the latter of which is an ordinary function in $t$ regardless of what the nature of $\mathrm{I}(t,u+\tau)$ is. Once we integrate $\mathrm{I}(t,u+\tau) \lambda^\tau (u)$ with respect to $u$ over the singleton $\{0\}$ in the case $t = \tau$, integration will only pick up the generalised function part of $\mathrm{I}(\tau,u+\tau)$, i.e., $\delta\big(\tau-(u+\tau)\big)= \delta(u)$, producing the term $\lambda^\tau(0)$, as intended. 
\end{rem}

\begin{rem}\label{rem:rigorous}
When applying the Leibniz integral rule in \eqref{eq:Leibniz}, we have not attempted to verify its assumptions. In fact, doing so would be difficult since we do not know a priori that cumulative incidence is differentiable with respect to time. Proving its differentiability from first principles using Lebesgue's dominated convergence theorem would similarly be difficult since it is not straightforward to derive sufficiently sharp a priori estimates for the increments of $t \mapsto \mathrm{CI}(t,\tau)$.
However, there is an alternative way of proving \eqref{eq:inc-eq} and \eqref{eq:inc-eq-2} rigorously, which can be outlined as follows. We first treat these equations as an educated guess and show they have a (unique) solution. We can then show that the time-integral of the solution satisfies the equation \eqref{eq:cinc-eq} for cumulative incidence. Finally, it is straightforward to prove uniqueness of solutions for \eqref{eq:cinc-eq} using Gr\"onwall's lemma (cf.\ Appendix~\ref{sec:Appendix2}), which then lets us conclude that the time-derivative of cumulative incidence indeed follows \eqref{eq:inc-eq-2}.  We will elaborate on the remaining mathematical details of this argument, including rigorous treatment of the Dirac delta function as a generalised function, in a separate paper. 
\end{rem}

\begin{exm}[\textbf{Incidence for the Bellman--Harris process and Poisson process model}]
Under the aforementioned assumptions, equations \eqref{eq:inc-eq} and \eqref{eq:inc-eq-2} read as
\begin{align}
\mathrm{I}(t,\tau) &= R(t)g^\tau(t-\tau)  + \int_{(0,t-\tau)} \mathrm{I}(t,u+\tau) R(u+\tau) g^\tau(u) \ud u, & t > \tau \geq 0,\\
\mathrm{I}(t,\tau) &= \delta(t-\tau)+\int_{[0,t-\tau]} \mathrm{I}(t,u+\tau) R(u+\tau) g^\tau(u) \ud u, & t \geq \tau \geq 0,\notag
\end{align}
respectively, for the Bellman--Harris process of Examples \ref{exm:BH1} and \ref{exm:BH2}, and as
\begin{align*}
\mathrm{I}(t,\tau) &= \rho(t)k(t-\tau)\overline{G}^\tau(t-\tau)  + \int_{(0,t-\tau)} \mathrm{I}(t,u+\tau) \rho(u+\tau)k(u) \overline{G}^\tau(u) \ud u, & t > \tau \geq 0,\\
\mathrm{I}(t,\tau) &= \delta(t-\tau)+\int_{[0,t-\tau]} \mathrm{I}(t,u+\tau) \rho(u+\tau)k(u) \overline{G}^\tau(u) \ud u, & t \geq \tau \geq 0,
\end{align*}
respectively, for the Poisson process model of Examples \ref{exm:Poisson} and \ref{exm:Poisson-2}.
\end{exm}

\subsection{Consistency with back-calculation}\label{sec:backcalculation}

Back-calculation is a standard  method to recover prevalence from incidence by convolving the survival function of the generation interval with incidence \cite{Brookmeyer1988-ag,Crump2015-sc}. We will now show that the equations we have obtained for prevalence and incidence are consistent with the back-calculation relationship under the assumption \eqref{eq:lambda-dens} and the additional assumption that the CDF $G^\tau(\cdot)$ does not depend on the infection time $\tau$, in which case we write $G(\cdot)$ and $\overline{G}(\cdot)$ \label{int:G-and-Gbar} in lieu of $G^\tau(\cdot)$ and $\overline{G}^\tau(\cdot)$, respectively.

Let $f$ and $\tilde{f}$ be two functions, one of which may be a generalised function, such that $f(t) = 0$ for any $t <0$ and $\tilde{f}(t) = 0$ for any $t < \tau$. Their convolution can be expressed as
\begin{equation*}
(f * \tilde{f})(t) \eqdefl \int_{[\tau,t]} f(t-s) \tilde{f}(s) \ud s
\end{equation*}
for any $t \geq \tau$ and equals zero otherwise. 
We proceed now to show that the back-calculation relationship
\begin{equation}\label{eq:back-calc}
\big(\overline{G} * \mathrm{I}(\,\cdot\,,\tau)\big)(t) = \mathrm{Pr}(t,\tau), \quad t \geq \tau \geq 0,
\end{equation}
holds, with the convention $\overline{G}(t) \eqdefl 0$ for any $t<0$.
Starting from \eqref{eq:inc-eq-2}, we have
\begin{equation}\label{eq:back-calc-2}
\overline{G} * \mathrm{I}(\,\cdot\,,\tau) = \overline{G} * \delta(\,\cdot\,-\tau) + \overline{G} * \int_{[0,\,\cdot\,-\tau]} \mathrm{I}(\,\cdot\,,u+\tau) \lambda^\tau (u) \ud u,
\end{equation}
where the first term on the right-hand side can be computed as
\begin{equation}\label{eq:back-c-1}
\big(\overline{G} * \delta(\,\cdot\,-\tau)\big)(t) = \int_{\R} \overline{G}(t-s) \delta(s-\tau) \ud s = \overline{G}(t-\tau), \quad t \geq \tau.
\end{equation}
The second term on the right-hand side of \eqref{eq:back-calc-2} vanishes for any argument $t \leq \tau$, so it suffices to consider $t > \tau$. In this case, switching the order of integration, we obtain
\begin{equation}\label{eq:back-c-2}
\begin{aligned}
\bigg(\overline{G} * \int_{[0,\,\cdot\,-\tau]} \mathrm{I}(\,\cdot\,,u+\tau) \lambda^\tau (u) \ud u\bigg)(t) 
& = \int_{[\tau,t]} \overline{G}(t-s) \int_{[0,s-\tau]} \mathrm{I}(s,u+\tau) \lambda^\tau (u) \ud u \, \ud s \\
& = \int_{[0,t-\tau]} \int_{[u+\tau,t]} \overline{G}(t-s) \mathrm{I}(s,u+\tau) \ud s \, \lambda^\tau (u)  \ud u \\
& = \int_0^{t-\tau} \big(\overline{G} * \mathrm{I}(\, \cdot \,,u+\tau)\big)(t)\lambda^\tau (u)  \ud u.
\end{aligned}
\end{equation}
Combining \eqref{eq:back-c-1} and \eqref{eq:back-c-2}, we have altogether
\begin{equation*}
\big(\overline{G} * \mathrm{I}(\,\cdot\,,\tau)\big)(t) = \overline{G}(t-\tau) + \int_0^{t-\tau} \big(\overline{G} * \mathrm{I}(\, \cdot \,,u+\tau)\big) (t) \lambda^\tau (u)  \ud u, \quad t \geq \tau \geq 0. 
\end{equation*}
Matching this with equation \eqref{eq:prev-eq} under the assumption \eqref{eq:lambda-dens}, we deduce
\begin{equation*}
\big|\big(\overline{G} * \mathrm{I}(\,\cdot\,,\tau)\big)(t) - \mathrm{Pr}(t,\tau)\big| \leq \int_0^{t-\tau} \big|\big(\overline{G} * \mathrm{I}(\, \cdot \,,u+\tau)\big) (t) -\mathrm{Pr}(t,u+\tau)\big| \lambda^\tau (u)  \ud u.
\end{equation*}
By an application of Gr\"onwall's inequality, as outlined in Appendix \ref{sec:Appendix2}, we can finally conclude that the back-calculation relationship \eqref{eq:back-calc} holds.

\begin{rem}[\textbf{Modelling HIV incidence from prevalence}]
HIV is an example of a disease where, due to long incubation times, routine surveillance generally returns prevalence --- not incidence \cite{Eaton2014}. However, what is of interest to policy makers is incidence, not prevalence \cite{Brown2014-rv}. Common approaches all make use of the back-calculation relationship through convolving a latent function for incidence with the survival function $\overline{G}(\cdot)$ \cite{Brown2014-rv,Nishiura2004-cp,Salomon_undated-ho}. Our argument above shows that there is no need to model incidence as a latent function, rather one can fit $\rho(\cdot)$ or $R(\cdot)$ directly to prevalence data using the prevalence integral equation for $\mathrm{Pr}(t,\tau)$, after which $\mathrm{I}(t,\tau)$ can be computed directly without need for a latent incidence function. This relationship therefore can help facilitate simpler or more pragmatic modelling choices.
\end{rem}

\subsection{Consistency with a common renewal equation model for incidence}\label{sec:consistency}

The key difference between our newly derived integral equations and the common renewal equation used \cite{Cori2013,Fraser2007,Nouvellet2018} is the inclusion of the parameter $\tau$ that initially arises due to the timing of the index case. The inclusion of $\tau$ means that we need to work with $\mathrm{I}(t,\tau)$, not simply $\mathrm{I}(t)$, and also gives rise to terms outside of the integral depending on whether one is interested in incidence, cumulative incidence or prevalence.

As in Section \ref{sec:backcalculation}, we assume that $G^\tau(\cdot)$ does not depend on $\tau$, i.e., we work with $G(\cdot)$, and we moreover assume that $G(\cdot)$ has a PDF $g(\cdot)$.\label{int:g}
In this context, when extended to accommodate the general initial infection time $\tau$, the common renewal equation for incidence is tantamount to the integral equation
\begin{equation}\label{eq:renewal}
\mathrm{I}_{\mathrm{Ren}}(t,\tau) = \delta(t-\tau) + R(t) \int_{[0,t-\tau]} \mathrm{I}_{\mathrm{Ren}}(t-u,\tau) g(u) \ud u, \quad t \geq \tau.
\end{equation}
We show that the renewal equation \eqref{eq:renewal} in fact agrees with the integral equation \eqref{eq:inc-eq-2} in the Bellman--Harris case, that is,
\begin{equation*}
\mathrm{I}(t,\tau)=\mathrm{I}_{\mathrm{Ren}}(t,\tau), \quad t \geq \tau \geq 0.
\end{equation*}
While we focus on the Bellman--Harris process (Examples \ref{exm:BH1} and \ref{exm:BH2}) here for notational simplicity, the argument also applies to the Poisson process model (Examples \ref{exm:Poisson} and \ref{exm:Poisson-2}) simply by replacing $R(\cdot)$ with $\rho(\cdot)$ and $g(\cdot)$ with $k(\cdot) \overline{G}(\cdot)$, respectively, throughout. 

To this end, we first introduce
\begin{equation}\label{eq:IcRen-def}
J(t,\tau) \eqdefl R(t) \int_{[0,t-\tau]} \mathrm{I}_{\mathrm{Ren}}(t-u,\tau) g(u) \ud u, \quad t \geq \tau \geq 0,   
\end{equation}
so that, given \eqref{eq:renewal},
\begin{equation}\label{eq:delta-removed}
\mathrm{I}_{\mathrm{Ren}}(t,\tau) = \delta(t-\tau) + J(t,\tau).    
\end{equation}
Applying \eqref{eq:delta-removed} to the integrand in \eqref{eq:IcRen-def} yields
\begin{align}
J(t,\tau) & = R(t) \int_{[0,t-\tau]} \big(\delta(t-u-\tau) + J(t-u,\tau)\big) g(u) \ud u \notag \\
& = R(t) g(t-\tau) + R(t) \int_0^{t-\tau} J(t-u,\tau) g(u) \ud u.\label{eq:int-eq-J}
\end{align}

Additionally, we introduce
\begin{equation}\label{eq:delta-removed-2}
\widetilde{J}(t,\tau) \eqdefl \int_{[0,t-\tau]} \mathrm{I}_{\mathrm{Ren}}(t,u+\tau) R(u+\tau) g(u) \ud u, \quad t \geq \tau \geq 0.    
\end{equation}
Subsequently, by applying \eqref{eq:renewal} to the integrand in \eqref{eq:delta-removed-2} and switching the order of integration, we obtain
\begin{align}
\widetilde{J}(t,\tau) & = \int_{[0,t-\tau]} \bigg(\delta(t-\tau-u) + R(t) \int_{[0,t-\tau-u]} \mathrm{I}_{\mathrm{Ren}}(t-s,u+\tau) g(s) \ud s  \bigg) R(u+\tau) g(u) \ud u \notag \\
& = R(t)g(t-\tau) + R(t) \int_{[0,t-\tau]}   \int_{[0,t-\tau-u]} \mathrm{I}_{\mathrm{Ren}}(t-s,u+\tau) R(u+\tau) g(u) g(s) \ud s  \ud u \notag \\
& = R(t)g(t-\tau) + R(t) \int_{[0,t-\tau]} \int_{[0,t-\tau-s]} \mathrm{I}_{\mathrm{Ren}}(t-s,u+\tau) R(u+\tau) g(u) \ud u \, g(s) \ud s \notag \\
& = R(t)g(t-\tau) + R(t) \int_0^{t-\tau} \widetilde{J}(t-s,\tau) g(s) \ud s.\label{eq:int-eq-J-tilde}
\end{align}
The integral equations \eqref{eq:int-eq-J} and \eqref{eq:int-eq-J-tilde} then imply the bound
\begin{equation*}
\big|J(t,\tau)-\widetilde{J}(t,\tau)\big|   \leq R(t) \int_0^{t-\tau} \big|J(t-u,\tau)-\widetilde{J}(t-u,\tau)\big| g(u) \ud u,
\end{equation*}
and applying Gr\"onwall's inequality as outlined in Appendix \ref{sec:Appendix2} we deduce that
\begin{equation}\label{eq:J-eq}
    J(t,\tau) = \widetilde{J}(t,\tau), \quad t \geq \tau \geq 0.
\end{equation}

Finally, by \eqref{eq:delta-removed} and \eqref{eq:J-eq},
\begin{equation*}
\mathrm{I}_{\mathrm{Ren}}(t,\tau) = \delta(t-\tau) + \widetilde{J}(t,\tau) = \delta(t-\tau) + \int_{[0,t-\tau]} \mathrm{I}_{\mathrm{Ren}}(t,u+\tau) R(u+\tau) g(u) \ud u.   
\end{equation*}
Given \eqref{eq:inc-eq-2} in the Bellman--Harris case, we then have the bound
\begin{equation*}
|\mathrm{I}(t,\tau) - \mathrm{I}_{\mathrm{Ren}}(t,\tau)| \leq \int_0^{t-\tau} |\mathrm{I}(t,u+\tau) - \mathrm{I}_{\mathrm{Ren}}(t,u+\tau)| R(u+\tau) g(u) \ud u  \end{equation*}
for any $t \geq \tau \geq 0$. Applying the result in Appendix \ref{sec:Appendix2} again we conclude that, indeed, $\mathrm{I}(t,\tau) = \mathrm{I}_{\mathrm{Ren}}(t,\tau)$ holds for any $t \geq \tau \geq 0$.

\begin{rem}[\textbf{Equivalence does not extend beyond incidence}]
In the case of prevalence or cumulative incidence, the equivalence between the common renewal equation and our newly derived integral equations is broken. This is easy to see by examining the derivations leading to \eqref{eq:int-eq-J} and \eqref{eq:int-eq-J-tilde}. If we considered cumulative incidence, for example, a constant one instead of a Dirac delta function would appear and the leading terms in \eqref{eq:int-eq-J} and \eqref{eq:int-eq-J-tilde} would no longer agree, rendering the rest of the argument impossible to carry through. This illustrates why the common renewal equation is a special case of our integral equations only when the index case is infected at time $\tau= 0$ and when considering incidence. Simpler renewal equations that do not involve varying dependence on the parameter $\tau$ for prevalence or cumulative incidence are not possible.
\end{rem}

\section{Numerical implementation and empirical application}

\subsection{Discretisation of integral equations}\label{ssec:discretisation}

The integral equations for cumulative incidence, prevalence and incidence under the assumption \eqref{eq:lambda-dens} are all special cases of a generic equation
\begin{equation}\label{eq:integral-equation}
f(t,\tau) = h(t,\tau) + \int_{0}^{t-\tau} f(t,u+\tau) \lambda^{\tau}(u) \ud u, \quad t \geq \tau \geq 0,
\end{equation}
with the choices
\begin{equation}\label{eq:h-spec}
h(t,\tau) \eqdefl \begin{cases}
1, & f = \mathrm{CI}, \\
\overline{G}^\tau(t-\tau), & f = \mathrm{Pr}, \\
\lambda^{\tau}(t-\tau), & f = \mathrm{I} \text{ (for $t > \tau$).}
\end{cases}
\end{equation}
Recall that for the Bellman--Harris process of Examples \ref{exm:BH1} and \ref{exm:BH2},  we substitute $\lambda^\tau(u) \ud u \eqdefl R(u+\tau)  g^\tau(u) \ud u$ and for the Poisson process model of Examples \ref{exm:Poisson} and \ref{exm:Poisson-2}, $\lambda^\tau(u) \ud u \eqdefl  \rho(u+\tau)k(u) \overline{G}^\tau (u)  \ud u$.

A key hurdle in solving the equation \eqref{eq:integral-equation} is that on the right-hand side, we get $f(t,u)$ for $\tau\leq u\leq t$ and not $f(u,\tau)$ for $\tau \leq u \leq t$. What this means is that in order to solve $f(t,0)$ for $t \geq 0$, say, we need to actually solve $f(t,\tau)$ for any pair $(t,\tau)$ such that $t \geq \tau \geq 0$. This is in fact why we left the initial infection time $\tau$ as a free parameter. (Alternatively, we could view \eqref{eq:integral-equation} as a system of coupled integral equations, indexed by $\tau$, that need to be solved simultaneously.)

\begin{algorithm}[p]
\caption{Discretisation of integral equations}\label{alg:simple}
\begin{algorithmic}[1]
\Require functions $h(\cdot,\cdot)$ and $\lambda^{\cdot}(\cdot)$
\Require step size $\Delta>0$
\Require number of time steps $N \in \N$
\For{$n = 0,\ldots,N$}
\For{$i = 0,\ldots,n$}
\If{$i=0$}
\State $\widehat{f}_{n\Delta}(i\Delta) \gets h(n\Delta,n\Delta)$
\Else
\State $\widehat{f}_{n\Delta}(i\Delta) \gets {\displaystyle h\big(n\Delta,(n-i)\Delta\big) + \sum_{j=1}^i \widehat{f}_{n\Delta}\big((i-j)\Delta\big) \lambda^{(n-i)\Delta} (j\Delta) \Delta}$
\EndIf
\EndFor
\EndFor
\State\Return $\widehat{f}_{n\Delta}(n\Delta)$ ($\approx f(n\Delta,0)$) for any $n=0,\ldots,N$
\end{algorithmic}
\end{algorithm}

\begin{algorithm}[p]
\caption{Discretisation of integral equations, vectorised}\label{alg:vector}
\begin{algorithmic}[1]
\Require functions $h(\cdot,\cdot)$ and $\lambda^{\cdot}(\cdot)$
\Require step size $\Delta>0$
\Require number of time steps $N \in \N$
\State $H \gets \begin{bmatrix} h(0,0) & 0 & 0 & \hdots & 0\\
h(\Delta,\Delta) & h(\Delta,0) & 0 & \hdots & 0\\
h(2\Delta,2\Delta) & h(2\Delta,\Delta) & h(2\Delta,0) & \ddots & \vdots \\
\vdots  &  \vdots & \vdots & \ddots & 0 \\
h(N\Delta,N\Delta) & h(N\Delta,(N-1)\Delta) & h(N\Delta,(N-2)\Delta) & \hdots & h(N\Delta,0)
\end{bmatrix}$
\State $L \gets \Delta \begin{bmatrix} \lambda^{0}(N\Delta) & \lambda^{0}((N-1)\Delta) & \lambda^{0}((N-2)\Delta) & \hdots & \lambda^0 (\Delta)\\
0 & \lambda^{\Delta}((N-1)\Delta) & \lambda^{\Delta}((N-2)\Delta) & \hdots & \lambda^{\Delta} (\Delta)\\
0 & 0 & \lambda^{2\Delta}((N-2)\Delta) & \hdots & \lambda^{2\Delta} (\Delta) \\
\vdots  &  \vdots & \ddots & \ddots & \vdots \\
0 & 0 & \hdots & 0 & \lambda^{(N-1)\Delta} (\Delta)
\end{bmatrix}$
\State $F \gets \text{empty $(N+1) \times (N+1)$ matrix}$
\State $F[1:(N+1),1] \gets H[1:(N+1),1]$
\For{$i = 1,\ldots,N$}
\State $B \gets F[(i+1):(N+1),1:i]\odot L[1:(N-i+1),(N-i+1):N]$
\State $F[(i+1):(N+1),i+1] \gets H[(i+1):(N+1),i+1] + \mathrm{RowSum}(B)$
\EndFor
\State\Return $\mathrm{diag}(F) = \big(\widehat{f}_{0}(0), \widehat{f}_{\Delta}(\Delta),\ldots, \widehat{f}_{N\Delta}(N\Delta) \big)$
\end{algorithmic}
\end{algorithm}

Solving the equation \eqref{eq:integral-equation} numerically is greatly facilitated if we introduce the auxiliary quantity $f_c(t) \eqdefl f(c,c-t)$\label{int:fc1} for any $c \geq t \geq 0$. From \eqref{eq:integral-equation} we can deduce that, for fixed $c \geq 0$, the \emph{single-argument} function $f_c(\cdot)$ is governed by the renewal-like integral equation
\begin{equation*}
f_c(t) = h(c,c-t) + \int_0^t f_c(t-u) \lambda^{c-t}(u) \ud u, \quad c \geq t \geq 0.
\end{equation*}
We then recover $f(t,0)$ for $t \geq 0$ via $f(t,0) = f_t(t)$. In practice, we are interested in solving $f(t,0)$ discretely for $t = 0,\Delta,\ldots,N\Delta$ for some $N \in \N$ and $\Delta >0$.\label{int:N-and-Delta} To this end, we approximate $f_{n\Delta}(\cdot)$ recursively by\label{int:fhat}
\begin{equation*}
\widehat{f}_{n\Delta}(i\Delta) \eqdefl \begin{cases} f_{n\Delta}(0) = f(n\Delta,n\Delta) = h(n\Delta,n\Delta), & i = 0, \\
{\displaystyle h\big(n\Delta,(n-i)\Delta\big) + \sum_{j=1}^i \widehat{f}_{n\Delta}\big((i-j)\Delta\big) \lambda^{(n-i)\Delta} (j\Delta) \Delta,} & i = 1,\ldots,n,
\end{cases}
\end{equation*}
for any $n = 0,\ldots,N$. For clarity, we present the entire procedure in pseudo-code in Algorithm~\ref{alg:simple}.

\begin{exm}\label{exm:lambda}
Concretely, in the Bellman--Harris case, we set 
\begin{equation*}
\lambda^{(n-i)\Delta} (j\Delta) \eqdefl R\left((n-i+j)\Delta\right)g^{(n-i)\Delta}(j\Delta),
\end{equation*}
while in the case of the  Poisson process model,
\begin{equation*}
\lambda^{(n-i)\Delta} (j\Delta) \eqdefl \rho\left((n-i+j)\Delta\right)k(j\Delta)\overline{G}^{(n-i)\Delta}(j\Delta).
\end{equation*}
A simplified version of the algorithm for cumulative incidence in the Bellman--Harris case is found in Appendix \ref{sec:Appendix1}.
\end{exm}

In practice, the double for-loop in Algorithm \ref{alg:simple} may lead to computational inefficiency when $N$ is large and an interpreted language is used, so it is useful to refine it by vectorisation. To this end, for an $m \times n$ matrix $A$ and $1 \leq s \leq m$ and $1 \leq t \leq n$, we denote by $A[s,t]$ the $s$-th row, $t$-th column element of $A$.
Moreover, for $1 \leq i \leq j \leq m$ and $1 \leq k \leq l \leq n$, we write $A[i:j,k:l]$ for the sub-matrix consisting of each element
$A[s,t]$ where $i \leq s \leq j$ and $k \leq t \leq l$. (If $i=j$, we simply write $i$ in lieu of $i:j$.) We also denote by $\odot$ element-wise (Hadamard) multiplication of matrices. The vectorised version of Algorithm \ref{alg:simple} is given as Algorithm \ref{alg:vector}. This matrix computation is possible by observing that all relevant values of the functions $h(\cdot,\cdot)$ and $\lambda^{\cdot}(\cdot)$ can be stored in the matrices $H$ and $L$, respectively.
 Algorithm \ref{alg:vector} can be further vectorised with respect to parameters to produce simultaneously discretisations for multiple parameter values.

Additional computational savings could be attained in Algorithm \ref{alg:vector} by observing that the top-left corner of the matrix $L$ typically contains very small values since $g^{\tau}(u)$ and $\overline{G}^{\tau}(u)$ are small with large $u$. Therefore the matrices $L$ and $F$ could in practice be truncated with a small error in the computation of $\mathrm{diag}(F)$.  

We illustrate the use of these algorithms in Figure \ref{fig:Figure1}, where we
compute prevalence using Algorithm \ref{alg:vector} and compare the results with statistical estimates of prevalence from a Monte Carlo simulation.
Python implementations of Algorithms \ref{alg:simple} and \ref{alg:vector}, including a version of the latter vectorised over parameters, are provided as fully documented Jupyter notebooks in: \url{https://github.com/mspakkanen/integral-equations}

\subsection{Bayesian inference on empirical data}\label{ssec:Empirical}

We perform Bayesian inference to estimate the time-varying case reproduction number $\mathcal{R}(t)$, as defined in Remark \ref{rem:case_reproduction_numer}, for historical incidence data for Influenza \cite{Frost1919-ji}, Measles \cite{Groendyke2010-wh}, SARS \cite{Lipsitch2003-gy}  and Smallpox \cite{Gani2001-nd} and for recent SARS-CoV-2 serological prevalence data in the United Kingdom \cite{Pouwels2021-ly}. 

\subsubsection*{Historical incidence data}

\begin{table}[p]
\begin{center}
\boxed{\begin{minipage}[c]{0.9\linewidth}
\begin{align*}
    \phi & \sim \text{Normal}^+(0,2)\\
    \sigma &\sim \text{Exponential(50)}\\
    \epsilon & \sim \text{Normal}(0,\sigma)\\
   \text{Bellman--Harris} & \phantom{=} \begin{cases}
     R(t) &= R(t-1) + \epsilon_t\\
    \mathcal{R}(t)&=\int_t^\infty R(u)g(u-t) du \\
    \mathrm{I}(t,\tau) &= \delta(t-\tau)  +  \int_0^{t-\tau} \mathrm{I}(t,u+\tau) R(u+\tau) g(u) \ud u   \\  
    \end{cases}
    \\
      y &\sim  \text{Negative Binomial}(\mathrm{I}(t,0),\phi)\label{eq:bayesianBH}
\end{align*} \vspace{-0.5em}
\end{minipage}} 
\end{center}
\caption{Hierarchical Bayesian model for estimating incidence for a Bellman--Harris process}\label{tab:Bayes}
\end{table}

\begin{figure}[p]
\begin{center}
\includegraphics[width=\textwidth]{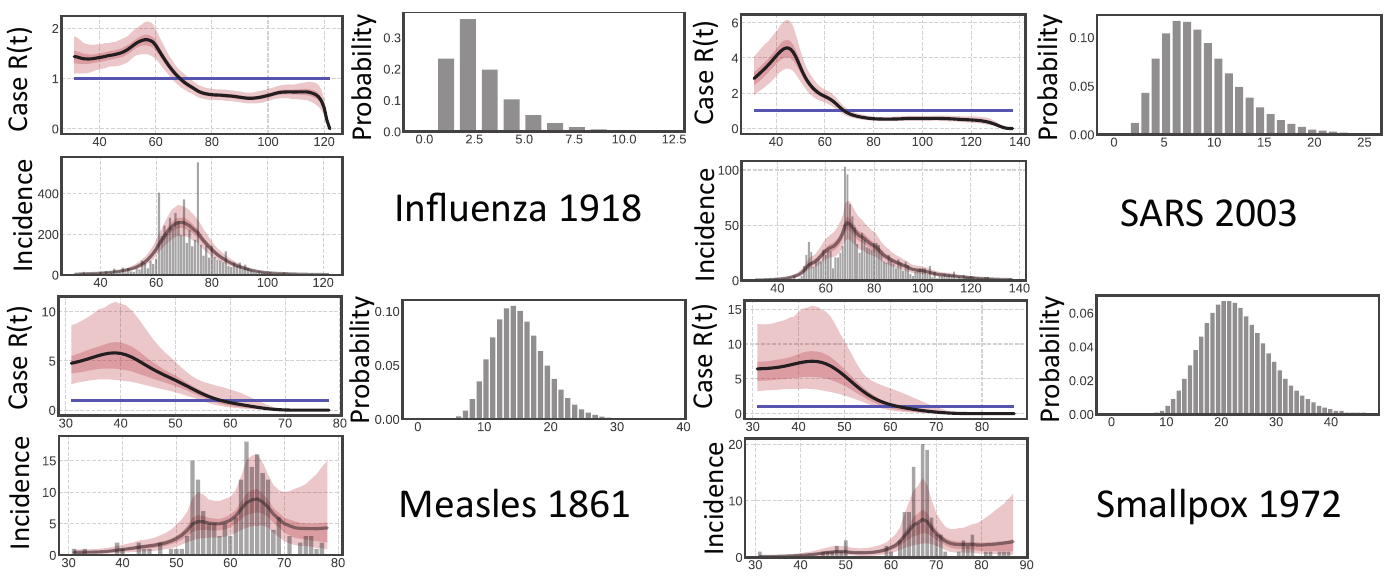}
\end{center}
\caption{Bayesian modelling of incidence for Influenza \cite{Frost1919-ji}, Measles \cite{Groendyke2010-wh}, SARS \cite{Lipsitch2003-gy} and Smallpox \cite{Gani2001-nd}. Plots show the case reproduction number $\mathcal{R}(t)$, the distribution $g(\cdot)$ in discretised form and incidence for the Bellman--Harris process.  Solid black lines in all plots are means, and the two red envelopes are the interquartile and $95\%$ credible intervals. The horizontal blue line indicates $\mathcal{R}=1$. The $x$-axis in all plots is time measured in days.}
\label{fig:Figure3}
\end{figure}

Historical incidence data for Influenza \cite{Frost1919-ji}, Measles \cite{Groendyke2010-wh}, SARS \cite{Lipsitch2003-gy} and Smallpox \cite{Gani2001-nd} have been extensively used in validating renewal equation frameworks \cite{Cori2013}. We fit an integral equation for the Bellman--Harris process. We work with $G^\tau(\cdot) = G(\cdot)$ that does not depend on $\tau$. As demonstrated in Section \ref{sec:consistency}, the corresponding integral equation agrees with the common renewal equation ubiquitously used in the modelling of incidence \cite{Cori2013}. 

We first introduce a probabilistic model for the function $R(\cdot)$ through a stochastic random walk process. To aid comparability to alternative methods \cite{Wallinga2004}, we transform $R(\cdot)$ to the case reproduction number $\mathcal{R}(t)$, which represents the average number of secondary cases arising from a primary case infected at time $t$, i.e., transmissibility after time $t$. In Table \ref{tab:Bayes}, the Negative Binomial likelihood is re-parameterised to the mean--variance formulation, $y$ is the observed count data (number of infections), $\phi$ is the overdispersion parameter and $\sigma$ is the random walk variance parameter. Therein, we write $\text{Normal}^+(0,a)$ for a normal distribution $\text{Normal}(0,a)$ constrained to the positive real axis. The observed count data and generation intervals were obtained from \cite{Cori_undated-pr,Cori2013}. The priors were selected to be weakly informative and were generally robust to change.

Algorithm \ref{alg:vector} was used to discretise and solve $t \mapsto \mathrm{I}(t,0)$ --- recall that $\tau$ is a parameter that is intrinsically involved in the solution of the integral equation, although we can ultimately restrict our attention to $t \mapsto \mathrm{I}(t,0)$ only, having assumed that the first infection occurs at time $\tau=0$. For all data sets, an arbitrary seeding period of $10$ days was used to correct for poor surveillance in the early epidemic. The seeding period was not included in the likelihood and we found our fits to be robust to different choices of seeding duration. Posterior sampling was performed using Hamiltonian Monte Carlo (1000 warmup/1000 sampling with multiple chains) in the Bayesian probabilistic programming language Numpyro \cite{bingham2019pyro,phan2019composable}. Posterior predictive checks were performed by examining R-hat and K-hat distributions. Figure \ref{fig:Figure3} shows the estimated case reproduction numbers $\mathcal{R}(t)$, which, as expected, match those previously estimated \cite{Cori2013}.

\subsubsection*{Serological prevalence data}

The ONS infection survey, is a weekly, household cross-sectional survey of blood samples which are used to test for the presence of COVID-19 antibodies, led by the Office for National Statistics (ONS) and the Department of Health and Social Care of the United Kingdom. At any point in time the ONS infection survey provides an estimate for the number of individuals currently infected with SARS-CoV-2, i.e., the prevalence of infection/positivity rates. Estimation of incidence from the ONS infection survey is done using a bespoke deconvolution approach, and estimating $R(t)$ or incidence directly from prevalence, to our knowledge, has not been attempted. 

\begin{table}[p]
\begin{center}
\boxed{\begin{minipage}[c]{0.9\linewidth}
\begin{align*}
    \phi & \sim \text{Normal}^+(0,2)\\
    \sigma &\sim \text{Exponential(50)}\\
    \epsilon & \sim \text{Normal}(0,\sigma)\\
   \text{Poisson process} & \phantom{=} \begin{cases}
     R(t) &= R(t-1) + \epsilon_t\\
    \mathcal{R}(t)&=\int_t^\infty R(u)g(u-t) du \\
    \mathrm{I}(t,\tau) &= \delta(t-\tau)  +  \int_0^{t-\tau} \mathrm{I}(t,u+\tau) \rho(u+\tau)k(u) g(u) \ud u   \\  
    \mathrm{Pr}(t,\tau) &= \overline{G}^\tau(t-\tau)  +  \int_0^{t-\tau} \mathrm{\overline{G}}(t,u+\tau) \rho(u+\tau)k(u) g(u) \ud u 
    \end{cases}
    \\
      y &\sim  \text{Negative Binomial}(\mathrm{Pr}(t,0),\phi)
\end{align*} \vspace{-0.5em}
\end{minipage}} 
\end{center}
\caption{Hierarchical Bayesian model for estimating prevalence for a Poisson process model}\label{tab:Bayes_sero}
\end{table}

\begin{figure}[p]
\includegraphics[width=\linewidth]{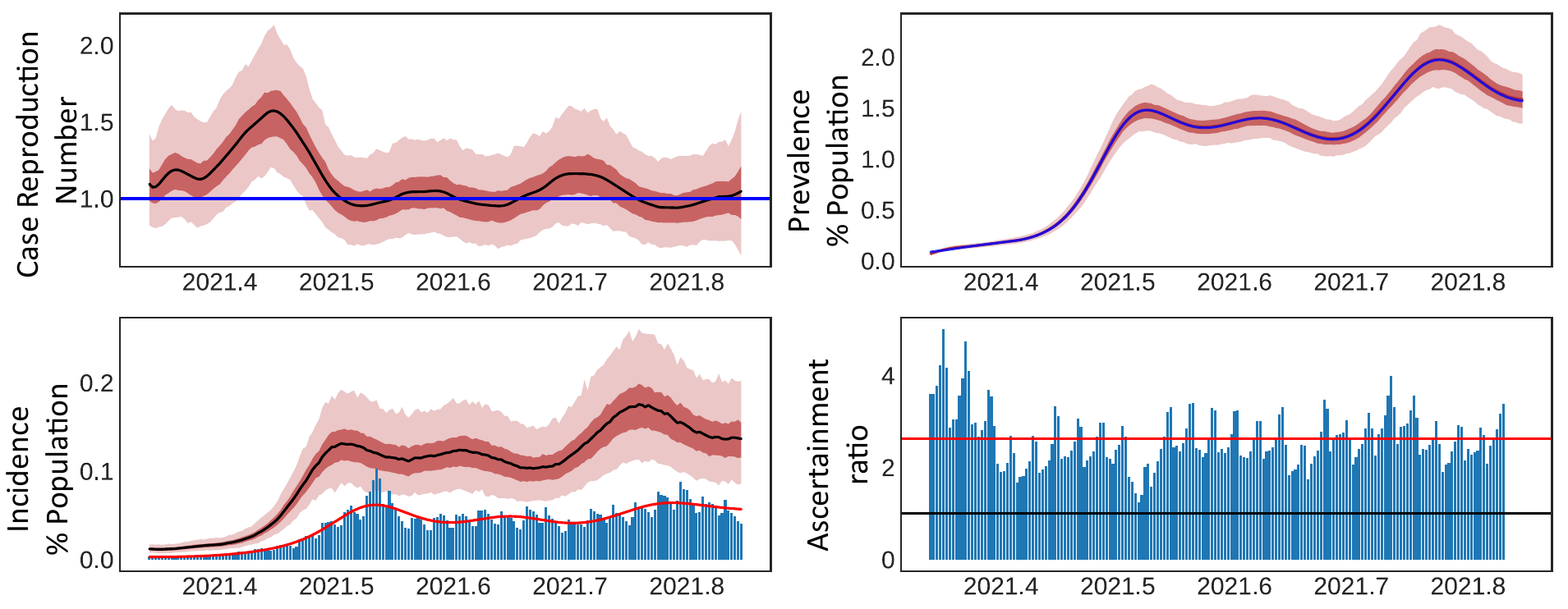}
\centering
\caption{Bayesian modelling of the ONS COVID-19 infection survey for prevalence. Top left show the case reproduction number $\mathcal{R}(t)$, top right prevalence, bottom left incidence and bottom right the ascertainment ratio (incidence/reported cases).  Solid black lines in all plots are means, and the two red envelopes are the interquartile and $95\%$ credible intervals. The horizontal blue line indicates $\mathcal{R}=1$. The $x$-axis in all plots is decimal calendar time. The ascertainment ratio in the bottom right is adjusted for the reporting delay between infections and cases, and this delay is estimated as the maximal lagged cross correlation.}
\label{fig:Figure4}
\end{figure}

We study estimates of prevalence from the ONS infection survey over the period 5th April 2021 to 15th November 2021. Our choice for this period arose from the requirement of wide spread, easily accessible SARS-CoV-2 PCR testing in the general population, which is required to ensure comparability between the ONS infection survey and reported case data (which we compare our estimates to). Prevalence estimates are reported weekly and we therefore use smoothing splines to interpolate these weekly estimates to daily estimates through a log linear generalised additive spline model \cite{Hastie2009-vl}. ONS infection survey results are generally reported on the Friday of any given week, with the sampling period covering Wednesday to Wednesday --- a period of 10 days. We therefore incorporate this observation lag by convolving daily prevalence with $\text{Normal}(10,0.3)$ distribution to adjust for these reporting lags and incorporate some uncertainty in this lag. We fit a Poisson process model, detailed in Table~\ref{tab:Bayes_sero}, to these lagged prevalence data assuming the infectiousness profile to be analogous to the generation time such that $k(\cdot)$ is given by the PDF of $\text{Gamma}(4.84,1.73)$ distribution \cite{Sharma2021-cs}. The CDF $G(\cdot)$ of the infection duration was assumed to follow the CDF of $\text{Normal}(10,1.5)$ distribution \cite{Wolfel2020-gk}. We did fit an aggregated likelihood where a daily Poisson process was aggregated to weekly averages, but found little difference in results.

The top left panel of Figure  \ref{fig:Figure4} shows the estimate case reproduction number $\mathcal{R}(t)$, which fluctuates around 1 over the period of study. The top right panel exhibits an excellent posterior fit to the daily smoothed ONS infection survey prevalence. Moreover, the bottom right panel shows infection incidence using the estimated $R(t)$ from fitting prevalence, and bars indicate the reported case data. Note we do not fit directly to the case data, but only to the prevalence as estimated by the ONS survey --- however including a second likelihood would be trivial to add. Lagging the time series and estimating the maximum cross correlation suggest a lag of approximately $7$ days between infections and reported cases --- a lag that is in line with previous studies \cite{Sharma2021-cs}. Finally, correcting for this lag between infections and cases, we see a reasonably stable (aside from weekly reporting cycles) infection ascertainment ratio (bottom right panel of Figure~\ref{fig:Figure4}) with a mean of approximately $2.5$, implying that for most of the study period there were 2.5 times more infections than reported and that this is relatively stable given testing policies over the period. This example demonstrates how our framework can fit prevalence directly without the need of deconvolution type approaches.

\section{Discussion}

Our primary goal in this paper is to bridge the worlds of individual-based models and mechanistic models to gain from the best of both. To this end, we began by choosing the most general branching process available --- the Crump--Mode--Jagers process \cite{Crump1968-rp,Crump1969-gc,Jagers1975-sd}. In the Crump--Mode--Jagers process, an epidemic is created at an individual level where, from a single infected individual, subsequent infections occur at random times according to their level of infectiousness. To our knowledge, for the first time, we generalise the Crump--Mode--Jagers process to allow for fully time-varying reproduction process for new infections. Indeed, rather than assuming the distribution of new infections to be constant (corresponding to a basic reproduction number) we allow it to change over time, which is essential in the modelling of real outbreaks \cite{Gostic2020-mm} beyond their early phase. We find that under this generalisation, a general integral equation arises from the Crump--Mode--Jagers process. Our framework also allows us to specify the dynamics of how new infections arise (in addition to them changing over time). Studying first the case where each infected individual produces all of their secondary cases, or ``offspring'', at the same random time, we recover the well known Bellman--Harris process \cite{Bellman1948}. Studying a more complex assumption where each infection can give rise to its offspring over the duration of its infection (an inhomogenous Poisson process) we derive a new integral equation, which to our knowledge, has not been previously presented. Remarkably, we find that despite the Poisson process model being much more complex than the simple Bellman--Harris assumption, the resultant integral equation has exactly the same form as the Bellman--Harris integral equation, only instead of the generation interval CDF, the survival probability is used. 

Through starting from a stochastic process, we are able to define prevalence, incidence and cumulative incidence as summary statistics (via moments) of an individual-based infection process. The benefit of defining these well known epidemiological quantities from a single stochastic process is that they are, by design, consistent with one another --- i.e., they are parameterised with the same generation interval and transmission rate (either $\rho(t)$ or $R(t)$). This allows practitioners to fit to prevalence for example, and easily recover incidence with no additional fitting. We mathematically show that this is the case and prove our equations for prevalence and incidence are consistent under the commonly used back-calculation technique in epidemiology \cite{Brookmeyer1988-ag}. Given ever increasing amount of infectious disease surveillance, being able to model prevalence and incidence simultaneously under the same process can greatly improve estimates of the rates of the reproduction number. A recent example is the COVID-19 pandemic where several countries collected high quality data on both cases (incidence) and serology (prevalence) \cite{Flaxman2020-lt}.  

We also show that the incidence integral equations we recover from the Bellman--Harris process and from the Poisson process model are in fact in agreement with the renewal equation commonly used in the modelling of incidence \cite{Cori2013}. Specifically, the common renewal equation is a special case of our incidence equations under the scenario where the first infection occurs at a specific, non-random, time. We also show that our equations are more general, and accommodate the modelling of prevalence, cumulative incidence, complex importation functions, and time-varying generation times \cite{Kimmel1983-jo}. The common renewal equation is computationally simpler as it does not involve the time $\tau$ of the first infection and simplifies the problem from two-dimensional to one-dimensional. We have however introduced an efficient algorithm which relies on straightforward matrix algebra to compute our more general integral equations. Given the ability of modern computers to perform matrix operations efficiently, we do not believe the computational overhead of our integral equations is meaningfully greater than that of the simple renewal equation. However, our integral equations allow for a far greater range of modelling choices with explicitly stated assumptions.

In this work, we have attempted to put the modelling of infectious diseases using renewal equations on firm mathematical ground. These mathematical foundations are broad enough to cover a variety of model specifications for transmission dynamics, and from them we can extract information about a wide range of relevant epidemiological quantities. In doing so, we have once again  made explicit the connection between branching processes \cite{Bellman1948,Crump1968-rp} and renewal equations. Explicit links between renewal equations and SEIR models \cite{Champredon2018-sg} and Hawkes processes \cite{Rizoiu2017} have been previously noted. It is likely other such relationships exist, and this is an interesting area of further study. Of additional interest is to use our framework to study the more complex L\'evy and Cox process models, which may produce renewal equations with even more realistic dynamics. Equally, recent frameworks \cite{Gomez-Rodriguez2010,Routledge2018EstimatingSetting} have extended the seminal work of \cite{Wallinga2004} to estimate case reproduction number on graphs --- connecting these two approaches is an interesting area of future research. Finally, our framework, and the vast majority of previous frameworks, only consider the mean integral equation and ignore the dynamics of higher-order moments. Using our framework, we can recover these moments from our stochastic process and formulate more accurate likelihoods for model fitting.

\subsection*{Acknowledgements}

S.B. and C.W acknowledge support from the MRC Centre for Global Infectious Disease Analysis (MR/R015600/1), jointly funded by the UK Medical Research Council (MRC) and the UK Foreign, Commonwealth \& Development Office (FCDO), under the MRC/FCDO Concordat agreement, and also part of the EDCTP2 programme supported by the European Union. S.B. acknowledges support from the Novo Nordisk Foundation via The Novo Nordisk Young Investigator Award (NNF20OC0059309), which also supports S.M. S.B. acknowledges support from the Danish National Research Foundation via a chair position. S.B. acknowledges support from The Eric and Wendy Schmidt Fund For Strategic Innovation via the Schmidt Polymath Award (G-22-63345). S.B. acknowledges support from the  National Institute for Health Research (NIHR) via the Health Protection Research Unit in Modelling and Health Economics. 

\printbibliography
\pagebreak
\appendix
\section{Discretising cumulative incidence under the Bellman--Harris process}\label{sec:Appendix1}

For the convenience of the reader, we present a simplified example how the integral equation for cumulative incidence under the Bellman--Harris process (Examples \ref{exm:BH1} and \ref{exm:BH2}) can be discretised using the methodology of Section \ref{ssec:discretisation}. While we consider, for the sake of concreteness, cumulative incidence, corresponding to $h(\cdot,\cdot)=1$, adapting the example to prevalence or incidence is straightforward by re-specifying $h(\cdot,\cdot)$ following \eqref{eq:h-spec}. We also assume that $g^\tau(\cdot) = g(\cdot)$ does not depend on $\tau$. 

To solve for cumulative incidence $t \mapsto \mathrm{CI}(t,0)$, we work with the equation
\begin{equation*}
f_c(t) = \underbrace{h(c,c-t)}_{=1} + \int_0^t f_c(t-u) \lambda^{c-t}(u) \ud u, \quad c \geq t \geq 0,
\end{equation*}
connected with cumulative incidence via $\mathrm{CI}(t,0) = f_t(t)$, $t \geq 0$. Suppose we choose $\Delta = 1$ as step size and wish to obtain an approximant $\widehat{f}_n(n)$ of $f_n(n) = \mathrm{CI}(n,0)$ for $n = 0,\ldots,N$. Then, we simply perform the recursive computation
\begin{equation}\label{eq:app-example}
\widehat{f}_{n}(i) \eqdefl \begin{cases} f_{n}(0) = f(n,0) = h(n,0)=1, & i = 0, \\
{\displaystyle 1 + \sum_{j=1}^i \widehat{f}_{n}\big(i-j\big) \lambda^{n-i} (j) ,} & i = 1,\ldots,n.
\end{cases}
\end{equation}
As pointed out in Example \ref{exm:lambda}, for the Bellman--Harris process, $\lambda^{n-i} (j) \eqdefl R(n-i+j)g(j)$. For any $n$, the computation in \eqref{eq:app-example} requires a for-loop over $i$, as shown in Algorithm \ref{alg:bh-ci}. In the course of the computation, we store the values of $\widehat{f}_\cdot(\cdot)$ into an $(N+1) \times (N+1)$ matrix $F$, the diagonal of which will contain the values $\widehat{f}_{0}(0), \widehat{f}_{1}(1),\ldots, \widehat{f}_{N}(N)$ we are ultimately interested in. 
\begin{algorithm}[H]
	\caption{Discretisation of integral equations, cumulative incidence under Bellman--Harris} \label{alg:bh-ci}
	\begin{algorithmic}[1]
	\Require number of time steps $N \in \N$
	\Require generation time PDF $g(n)$ at $n = 1,\ldots,N$
	\Require reproduction number $R(n)$ at $n = 1,\ldots,N$
	\State $F \gets \text{empty $(N+1) \times (N+1)$ matrix}$
	\State $F[1 : (N+1), 1] \gets 1$
    \For {$n=1,\ldots,N$}
        \For {$i=1,\ldots,n$} 
             \State $F[n+1,i+1] \gets 1+\sum_{j=1}^i R(n-i+j) F[n+1,i-j+1]g(j)$
        \EndFor
	\EndFor
	\State\Return $\mathrm{diag}(F) = \big(\widehat{f}_{0}(0), \widehat{f}_{1}(1),\ldots, \widehat{f}_{N}(N) \big)$
	\end{algorithmic} 
\end{algorithm}
\begin{rem}
For the Poisson process model (Examples \ref{exm:Poisson} and \ref{exm:Poisson-2}), the algorithm would be identical except that $g(\cdot)$ is replaced by $k(\cdot)\overline{G}(\cdot)$ and $R(\cdot)$ is replaced by $\rho(\cdot)$.
\end{rem}

\section{Two particular forms of Gr\"onwall's inequality}\label{sec:Appendix2}

Suppose that a non-negative, locally bounded function $(t,\tau) \mapsto f(t,\tau)$ satisfies
\begin{equation}\label{eq:gronwall-1}
f(t,\tau) \leq \int_0^{t-\tau} f(t,u+\tau) b(t,\tau,u) \ud u, \quad t \geq \tau \geq 0,    
\end{equation}
for some bounded non-negative function $(t,\tau,u) \mapsto b(t,\tau,u)$. Here, we clarify how we can then deduce that
\begin{equation}\label{eq:gronwall-2}
f(t,\tau) = 0  \quad \text{for any $t \geq \tau \geq 0$.}
\end{equation}
Writing $\overline{b} \eqdefl \sup_{t,\tau,u} b(t,\tau,u)$, the inequality \eqref{eq:gronwall-1} implies
\begin{equation*}
f(t,\tau) \leq \overline{b}\int_0^{t-\tau} f(t,u+\tau)  \ud u, \quad t \geq \tau \geq 0.    
\end{equation*}
As earlier, consider $f_c(t) \eqdefl f(c,c-t)$ \label{int:fc2} for any $c \geq t \geq 0$, so that
\begin{equation*}
f_c(t) \leq \overline{b} \int_0^t f(c,u+c-t) \ud u = \overline{b} \int_0^t f_c(t-u) \ud u = \overline{b} \int_0^t f_c(u) \ud u.   
\end{equation*}
A general version of Gr\"onwall's inequality \cite[Theorem 3.1(d)]{horvath-1996}, which does not require $f$ to be known to be continuous a priori, then implies that $f_c(t) = 0$ for any $c \geq t \geq 0$, from which \eqref{eq:gronwall-2} readily follows. 

Another case we need is where
\begin{equation*}
f(t,\tau) \leq \int_0^{t-\tau} f(t-u,\tau)b(t,\tau,u) \ud u, \quad t \geq \tau \geq 0.   
\end{equation*}
Then we have
\begin{equation*}
f(t,\tau) \leq \overline{b} \int_0^{t-\tau} f(t-u,\tau) \ud u \leq     \overline{b}  \int_0^{t} f(t-u,\tau) \ud u \leq \overline{b}  \int_0^{t} f(u,\tau) \ud u,
\end{equation*}
and \eqref{eq:gronwall-2} follows by applying again \cite[Theorem 3.1(d)]{horvath-1996}.

\newpage

\section{Glossary of main symbols}

\begin{tabular}{m{0.17\textwidth}  m{0.57\textwidth}  m{0.075\textwidth}  m{0.095\textwidth}}
\hline
 Symbol & Description & Type & Page(s) \\
 \hline
 $\mathbf{1}_A(\cdot)$ & Indicator of a set $A$ & Func & \pageref{int:ind}\\
 $B$  & Matrix related to Algorithm \ref{alg:vector} & Mat & \pageref{alg:vector},\ \pageref{alg:bh-ci} \\
 $\mathrm{CI}(\cdot,\tau)$ & Cumulative incidence (of an epidemic started at $\tau$) & Func & \pageref{int:cinc} \\
 $F$ & Matrix related to Algorithms \ref{alg:vector} and \ref{alg:bh-ci} & Mat & \pageref{alg:vector},\ \pageref{alg:bh-ci}\\
  $f(\cdot,\cdot)$  & Function satisfying an integral equation (or inequality) & Func & \pageref{f-int},\ \pageref{eq:integral-equation},\ \pageref{eq:gronwall-1}\\
 $f_c(\cdot)$ & Auxiliary function defined via $f_c(t) \eqdefl f(c,c-t)$ & Func & \pageref{int:fc1},\ \pageref{int:fc2}\\
 $\widehat{f}_c(\cdot)$ & Discretisation of $f_c(\cdot)$ via Algorithms \ref{alg:simple} and \ref{alg:vector} & Func & \pageref{int:fhat} \\
  $G(\cdot)$, $G^\tau(\cdot)$ & Cumulative distribution functions of $L^\tau$ & Func & \pageref{int:Gtau},\ \pageref{int:G-and-Gbar} \\
 $\overline{G}(\cdot)$, $\overline{G}^\tau(\cdot)$ & Survival functions of $L^\tau$ & Func & \pageref{int:Gtaubar},\ \pageref{int:G-and-Gbar} \\
 $g(\cdot)$, $g^\tau(\cdot)$ & Probability density functions of $L^\tau$ & Func & \pageref{int:gtau},\ \pageref{int:g}\\
 $H$ & Matrix related to Algorithm \ref{alg:vector} & Mat & \pageref{alg:vector}\\
 $h(\cdot,\cdot)$ & Generic function (related to an integral equation) & Func & \pageref{eq:integral-equation} \\
  $\mathrm{I}(\cdot,\tau)$ & Incidence (of an epidemic started at $\tau$) & Func & \pageref{int:inc} \\
 $\mathrm{I}(\cdot)$,\ $\mathrm{I}_{\mathrm{Ren}}(\cdot,\tau)$ & Incidence (defined via the common renewal equation) & Func & \pageref{int:inc-ren},\ \pageref{eq:renewal} \\
 $L$ & Matrix related to Algorithm \ref{alg:vector} & Mat & \pageref{alg:vector}\\
  $L^\tau$, $L^\tau_i$ & Infection lengths & RV & \pageref{int:Ltau}\\
  $k(\cdot)$ & Infectiousness profile (in the Poisson process model) & Func & \pageref{exm:Poisson}\\
    $N$ & Number of time steps & Const & \pageref{int:N-and-Delta}\\
 $N^\tau(\cdot)$, $N^\tau_i(\cdot)$ & Reproduction processes & SP & \pageref{int:Ntau}\\
  $\mathrm{Pr}(\cdot,\tau)$ & Prevalence (of an epidemic started at $\tau$) & Func & \pageref{int:prev} \\
 $R(\cdot)$ & Instantaneous reproduction number (in the Bellman--Harris process) & Func & \pageref{exm:BH1}\\
  $\mathcal{R}_{\mathrm{BH}}(\cdot)$, $\mathcal{R}_{\mathrm{Pois}}(\cdot)$ & Case reproduction numbers & Func & \pageref{rem:case_reproduction_numer}\\
   $Z(\cdot,\tau)$ & Branching process (counted via characteristic
   $\chi^\tau(\cdot)$) & SP & \pageref{int:Zproc} \\
    $\Delta$ & Discretisation step size & Const & \pageref{int:N-and-Delta}\\
     $\delta(\cdot)$ & Dirac delta function & Func & \pageref{int:delta} \\
      $\Lambda^{\tau}(\cdot)$ & Expectation of $N^\tau(\cdot)$ as a function of time & Func & \pageref{int:Lambda}\\
 $\lambda^\tau(\cdot)$ & Time-derivative of $\Lambda^{\tau}(\cdot)$ & Func & \pageref{int:lambda} \\
      $\xi(\cdot)$ & Number of infections generated at a given time (in the Bellman--Harris process) & SP & \pageref{exm:BH1}\\
       $\rho(\cdot)$ & Population-level variation in transmissibility (in the Poisson process model) & Func & \pageref{exm:Poisson} \\
 $\tau$ & Infection time of the index case & Const & \pageref{int:tau}\\
 $\tau_i$ & Infection time of a secondary case & RV & \pageref{int:taui}\\
  $\Phi(\cdot)$ & Homogeneous Poisson process & SP & \pageref{exm:Poisson}\\
 $\chi^\tau(\cdot)$, $\chi_i^\tau(\cdot)$ & Random characteristics & SP & \pageref{int:chitau}\\
\hline
\end{tabular}  \\  \\
\emph{Key:}  ``Const'' -- Constant, ``Func'' -- Function, ``Mat'' -- Matrix, ``RV'' -- Random variable, ``SP'' -- Stochastic process

\end{document}